\documentclass[fleqn,usenatbib,referee]{mnras}

\usepackage[T1]{fontenc}
\usepackage{ae,aecompl}
\usepackage[dvipsnames]{xcolor}

\usepackage{graphicx}	
\usepackage{amsmath}	
\usepackage{amssymb}	

\newcommand{\blap}[1]{\smash[b]{\begin{tabular}[t]{@{}c@{}}#1\end{tabular}}}

\title[Deep coorbitals of the Earth]{Long-term dynamical survival of deep Earth coorbitals}

\author[A. A. Christou et al.]{
Apostolos A. Christou,$^{1}$\thanks{E-mail: apostolos.christou@armagh.ac.uk (AAC)}
Nikolaos Georgakarakos$^{2,3}$
\\
$^{1}$Armagh Observatory and Planetarium, College Hill, Armagh BT61 9DG, Northern Ireland, UK\\
$^{2}$Division of Science, New York University Abu Dhabi, Abu Dhabi, UAE\\
$^{3}$Center for Astro, Particle and Planetary Physics ($\mbox{CAP}^3$), New York University Abu Dhabi, UAE
}
\date{Accepted 2021 July 26. Received 2021 July 26; in original form 2021 January 26}

\pubyear{2021}

\begin{document}
\label{firstpage}
\pagerange{\pageref{firstpage}--\pageref{lastpage}}
\maketitle

\begin{abstract}
We investigate the long-term dynamical survival of Earth co-orbital asteroids, focusing on near-circular, near-planar orbits which existing studies suggest are the most stable. Through numerical integration of test particles we show that about a quarter of an initial population can survive for at least 50\% of the age of the solar system with horseshoe particles being four to five times more likely to survive than L4/L5 Trojans. From the end state statistics we constrain the existence of planetesimal-sized objects originally in co-orbital libration, finding that typically $5^{-2}_{+7}$ such planetesimals and no more than $27^{-9}_{+30}$ (95\% confidence) could have been present. Our simulations also suggest that episodic variations in the terrestrial orbital eccentricity may have caused bulk escape of co-orbitals, though variations large enough ($>$0.01) to generate such episodes are statistically unlikely. We then consider the orbital evolution of co-orbital asteroids of sizes down to $D = 50$ m under the Yarkovsky effect and find that objects with $D$ $<$ 1 km should escape over 4 Gyr with the smallest asteroids escaping after 200 Myr. Further, we test whether Earth's co-orbital region may be populated by asteroids arriving via outward Yarkovsky drift, as conjectured by Zhou et al.~(A\&A, 622, A97, 2019). We find this is an inefficient process, as planetary close encounters rapidly scatter the orbits far from Earth's and towards the asteroid belt. Finally, we discuss how the destabilising action of Yarkovsky may be mitigated through spin state evolution or late collisional comminution of large parent asteroids.
\end{abstract}

\begin{keywords}
minor planets, asteroids: general -- planets and satellites, individual: Earth -- planets and satellites: dynamical evolution and stability -- celestial mechanics -- methods: numerical
\end{keywords}



\section{Introduction}
The existence of co-orbital companions of the Earth has been a long-standing question in planetary science \cite[see][for a review]{Malhotra2019}. These objects do not venture far from the Sun in the sky and ground-based surveys to-date have failed to directly detect them \cite[][and references therein]{Markwardt.et.al2020}. Although we now know of numerous coorbital companions of our planet, these are dynamically transient \citep{Christou2000a,Connors.et.al2004} and likely to be relatively recent arrivals from the same source regions that replenish the NEA population \citep{MoraisMorbidelli2002}. Despite these negative results, the interest to find the longest-lived or ``deep'' coorbitals remains strong because of the unique scientific value of such bodies as potential witnesses of the origin and early evolution of the Earth and the terrestrial planets \citep{Malhotra2019}.

Ground-based observational surveys, the most recent of which was the search for L5 Trojans with the DECam instrument \citep{Markwardt.et.al2020}, have now been complemented by {\it in situ} searches of the region around the Sun-Earth-asteroid $\mbox{L}_{5}$ equilibrium point, by the {\it OSIRIS-REx} and {\it Hayabusa II} spacecraft \citep{Cambioni.et.al2018,Yoshikawa.et.al2018}. The observational completeness achieved for these surveys, all of which yielded no Trojan detections, argues against the existence of Earth Trojans with absolute magnitude $H$$<$$15.5$ ($\sim$$3$ km across or larger for a visual albedo $p_{V}$=0.15) but allows for 100 or so objects with $H\simeq 20$, equivalent to sizes of up to a few hundred m across \citep{Cambioni.et.al2018,Markwardt.et.al2020}.

Lagging somewhat behind the observational effort has been our understanding of the dynamical stability of Earth co-orbitals - roughly speaking, those with semimajor axis $a$ between 0.99 and 1.01 au. While numerous studies have demonstrated stability for up to $10^{7}-10^{8}$ yr \cite[eg][]{TabachnikEvans2000a}, to-date there have been only a few attempts to directly quantify the survival of such objects over periods comparable to the age of the solar system. The latter studies were primarily motivated by the discovery, by the {\it WISE} space observatory, of the first Earth Trojan asteroid, 2010 $\mbox{TK}_{7}$ \citep{Connors.et.al2011}, albeit a transient one \citep{Dvorak.et.al2012} and of 2010 $\mbox{SO}_{16}$, a relatively long-lived but also transient Earth horseshoe asteroid \citep{ChristouAsher2011}.

\begin{figure}
\centering
\includegraphics[width=87mm,angle=0]{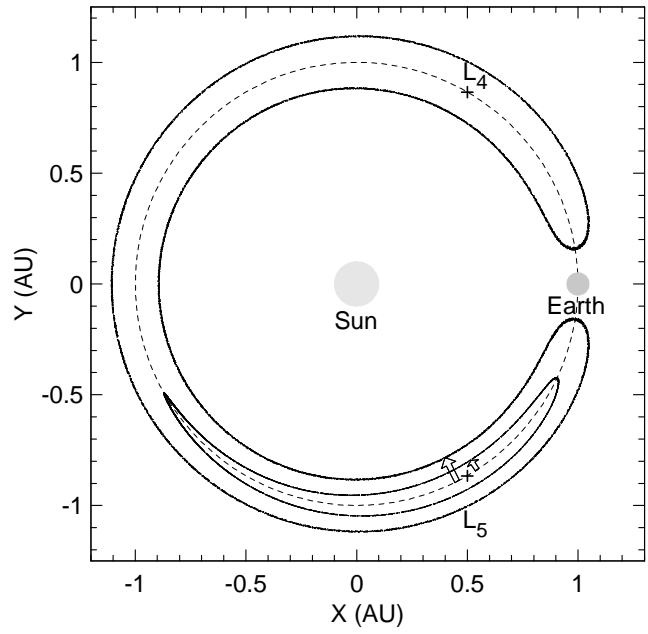}
\caption[Trajectories of Earth Trojan and horseshoe asteroids in a frame rotating with Earth's mean motion around the Sun. The arrows highlight the quantity $\Delta a$ used to parameterise the particle initial states.]{Trajectories of Earth Trojan and horseshoe asteroids in a frame rotating with Earth's mean motion around the Sun. The arrows highlight the quantity $\Delta a$ used to parameterise the particle initial states.}
\label{fig:trojan_horseshoe}
\end{figure}

\citet{Cuk.et.al2012} integrated particle ensembles within the coorbital region of the Earth for 700 Myr under the gravitational action of the planets. Two major outcomes of that study were (i) the robust identification of stable islands in phase space where particles dynamically persist for the full duration of the simulations, and (ii) that the longest-lived Earth coorbitals are not Trojans, where the motion is restricted in the vicinity of either the $\mbox{L}_{4}$ or $\mbox{L}_{5}$ equilibrium point, but horseshoes, following paths that encompass both $\mbox{L}_{4}$ and $\mbox{L}_{5}$ (Fig.~\ref{fig:trojan_horseshoe}). The most stable regions were found at low ($I<15^{\circ}$) inclination and, in addition, within a moderately stable region at $I > 25^{\circ}$ which is, however, efficiently cleared of particles after $10^{8}$ yr. A subsequent study by \citet{MarzariScholl2013} focused specifically on Trojans, mapping out the phase space using frequency map analysis. The authors found the most stable orbits to be those with $e \lesssim 0.1$, $I\lesssim50^{\circ}$ but unconstrained in the amplitude of the critical angle.

More recently, \citet{Zhou.et.al2019} produced a high-resolution stability map of Earth-coorbital phase space, building on earlier work by \citet{Dvorak.et.al2012} and using as proxy the number of lines in the frequency spectrum of the orbit history, the so-called Spectral Number. The topology of the stable domain agreed closely with the islands identified by \citet{Cuk.et.al2012} with horseshoe orbits found to be overall more stable than tadpole orbits. The authors moreover identified secular resonances that occupy the interstitial spaces between the different stable islands. Based on the resonance locations, they concluded that horseshoe orbits are destabilised by inclination-type resonances where the critical argument involves the longitude of the node. Tadpole orbits, on the other hand, are destabilised by eccentricity-type resonances involving the longitude of the perihelion as well as mixed-type resonances that involve both the nodes and the apses.

The non-gravitational Yarkovsky effect drives asteroid migration across the Main Belt and into the terrestrial planet region \citep{Bottke.et.al2006}. Yarkovsky, which generally modifies the keplerian semimajor axis $a$ of the orbit, has been shown to cause orbital migration and even escape of Trojans of Mars \citep{Cuk.et.al2015,Christou.et.al2020} and should, therefore, bear on the dynamical stability of Earth co-orbitals as well.
A simple observational argument for the long-term dynamical survival of Earth co-orbitals against Yarkovsky is provided by the existence of stable Mars co-orbitals \citep{Scholl.et.al2005}. This planet has two deep Trojans, 5261 Eureka and (121514) 1999 $\mbox{UJ}_{7}$, both with diameters $D \sim 2$ km \cite[][and references therein]{Christou.et.al2020}. Considering that, in the limit of a thermally ``relaxed'' asteroid \cite[][see also \citet{Bottke.et.al2006} and references therein]{Xu.et.al2020}, the radial drift rate would vary approximately as $\dot{a} \propto a^{-2} D^{-1}$, a rough size estimate for an asteroid at 1 au that would suffer the same amount of Yarkovsky orbital drift $\dot{a}$ is 2 $\times$ $(1\mbox{ au}/1.5\mbox{ au})^{-2}$ $\gtrsim$ 4 km.

\citet{MarzariScholl2013} were first to investigate the effect of Yarkovsky on the longevity of Earth co-orbitals. They carried out several-Gyr numerical simulations of Earth Trojans, finding these to survive for about a Gyr with the result being only weakly dependent on size. \citet{Zhou.et.al2019} repeated this type of study over a wider swath of phase space for both Trojan and horseshoe modes, integrating each of $\sim$1000 of the longest-lived particles in their gravity-only runs under both positive and negative Yarkovsky drift rates $\dot{a}$ in the range $|0.4 - 4| \times 10^{-3}$ au $\mbox{Myr}^{-1}$ for 1 Gyr. By extrapolating the observed loss rate of particles from the resonance, they concluded that long-lived particles in the gravitational problem would be efficienty cleared under the Yarkovsky effect so that $< 1$\% of the original population would survive over the age of the solar system. It was also found that orbits with $\dot{a}>0$ are somewhat more long-lived than those with $\dot{a}<0$, the implication being that Yarkovsky evolution within the Earth co-orbital region should favour prograde over retrograde rotators. They further speculated that Earth's coorbital region may be predominantly populated by prograde-rotating asteroids with orbits originally interior to the Earth's.

The present work serves a number of objectives. Firstly, the relative persistence of horseshoes over Trojans differs from other solar system locales where long-lived populations of co-orbitals are observed - namely at Mars, Jupiter \& Neptune. The implications for observational searches are probably not yet appreciated and additional studies that highlight this feature are timely. Secondly, \citet{Christou.et.al2020} recently introduced a paradigm of the extant Trojan population at Mars as the product of ongoing creation and escape of asteroids driven by thermal forces. If asteroids have been occupying Earth's coorbital region to the present day, either they or, more likely, their offspring may exist among the known NEA population and it is desirable to better understand where these recently-escaped asteroids are most likely to appear. Additionally, the orbit evolution of Trojans of Mars has been found to depend on the planet's eccentricity history \citep{Cuk.et.al2015} and we want to find out to what extent this is also true for the Earth.

These objectives can be achieved by brute-force numerical simulations of test particles covering a significant fraction of the age of the solar system.

This paper is organised as follows: in the next Section we describe the different sets of simulations used to investigate the research objectives. Section 3 describes our results while Section 4 presents and discusses our principal conclusions.

\section{\label{sec:integrations}Simulation setup}
For all simulations in this work we used the ``hybrid'' symplectic scheme available within the {\it MERCURY} package \citep{Chambers1999} with an integration time step of 4 days. The scheme accurately handles close encounters between particles and planets by switching from mixed-variable symplectic to Bulirsch-Stoer state propagation within a certain distance from a planet. For all simulations reported here, this changeover threshold was set at 2 Hill radii. The solar system model in the simulations is strictly Newtonian and includes the eight major planets from Mercury to Neptune. Initial planetary state vectors were retrieved from the HORIZONS online ephemeris service \citep{Giorgini.et.al1996} at the J2000 epoch and both massive bodies and test particles were integrated to the same epoch before the start of the simulations. 

Our different simulation batches are summarised in Table~\ref{tab:sims}.
In our main batch, we integrated groups of 151 particles for each of six different values of the particle semimajor axis relative to the Earth $\Delta a = a-a_{\rm Earth}$, approximately equally-spaced in the range 0.001-0.01 au. We refer to these groups using the notation B[i] where $i$ is an integer from 1 to 6 with higher values indicating a higher initial $\Delta a$ value. Reference orbit elements for the test particles in our simulations are shown in Table~\ref{tab:ic}. Values of the argument of perihelion $\omega$ and longitude of ascending node $\Omega$ are those for the horseshoe asteroid 419624 (2010 $\mbox{SO}_{16}$) retrieved from HORIZONS at the epoch JD2456190.5. For the eccentricity and inclination, we have chosen $e=0.025$ and $I=5^{\circ}$ in order to place the particles in the stable regions identified in \citet{Cuk.et.al2012}. The value of the mean anomaly $M$ was chosen to satisfy $\lambda-\lambda_{\rm Earth}=300^{\circ}$, placing the particle near the $\mbox{L}_{5}$ Earth-Sun Lagrangian point. Starting conditions for individual test particles within each group were generated as 6-dimensional gaussian random variates from a covariance matrix, using the method described in \citet{Duddy.et.al2012}. For convenience, we used here the formal state covariance for the asteroid 2010 SO16 at JD 2456600.5 retrieved from the Near Earth Objects Dynamic site\footnote{https://newton.spacedys.com/neodys/}. The minimum and maximum eigenvalues of this covariance were $\sim 10^{-8}$ au and $\sim 6 \times 10^{-7}$ deg, corresponding to the semimajor axis and the mean longitude of the orbit.

For planar, circular orbits the $\Delta a$ quantity determines the type of libration (Fig.~\ref{fig:trojan_horseshoe}), either tadpole, where the critical angle librates around the $\mbox{L}_{5}$ (or $\mbox{L}_{4}$) equilibrium points, if
\begin{equation}
\label{eq:thb}
\Delta a < \Delta a_{\rm crit} =a_{\rm P} \sqrt{\frac{8}{3}\mu_{\rm P}}
\end{equation}
or horseshoe (libration over a wide arc that encompasses both $\mbox{L}_{4}$ and $\mbox{L}_{5}$) otherwise \citep{MurrayDermott1999}, with $\mu_{\rm P}$ and $a_{\rm P}$ being the planet-Sun mass ratio and the planet semimajor axis respectively. For the Earth, $\mu_{\rm P} \simeq 3$$\times$$10^{-6}$ and $\Delta a_{\rm crit}=2.82$ $\times$ $10^{-3}$ au, therefore particles in groups B1 and B2 simulate $\mbox{L}_{5}$ tadpoles while groups B3 to B6 simulate horseshoes. We note here that the term ``Trojan'' is historically taken to refer to tadpole libration, although \cite{Zhou.et.al2019} use the term to refer to either tadpoles or horseshoes. In this paper we adopt the tadpole$=$Trojan convention and use the term ``coorbital'' to generally refer to different modes of libration in the 1:1 resonance.

Another batch of simulations (B1E and B4E runs, Table~\ref{tab:sims}) was used to explore the dependence of the stability properties of the coorbitals on the Earth orbit evolution. To generate Earth orbit variants for this batch, the y-component of the Earth's initial cartesian position vector was changed by $\pm 2 \times 10^{-9}$ au. 

A further two simulation batches (B1Y and B4Y runs) included the non-gravitational Yarkovsky effect, here we have introduced the along-track component of the diurnal Yarkovsky acceleration from \citet{Farinella.et.al1998} as a user-defined force within MERCURY. The magnitude $\alpha_{\rm Y}$ of the acceleration vector for each clone has the form
\begin{equation}  \label{eq:yarko}
\alpha_{\rm Y}= f \alpha_{\rm Y,max}\mbox{,    }\alpha_{\rm Y,max}= C_{Y} \frac{\cos \zeta}{D} \\
\end{equation}
where $\zeta$ represents the rotation axis obliquity while the quantity $C_{Y}$ contains the dependence on the orbital state vector as well as certain bulk and surface properties of the body. In our simulations, $f$ was set to sample the magnitude of the Yarkovsky acceleration linearly and uniformly over the range $|\dot{a}|$ $<$ $17 \times 10^{-3}$ au $\mbox{Myr}^{-1}$, the upper bound corresponding to evaluating $\alpha_{\rm Y,max}$ with $D=50$ m, $P=2$ hr and $\zeta=0^{\circ}$. Other parameters are the bulk and surface densities, specific heat capacity $C$, thermal conductivity $K$, surface thermal emissivity $\epsilon$ and surface albedo $A$. For these we adopted the same values as in \citet{Christou2013}, namely a bulk density of 1 g $\mbox{cm}^{-3}$ and equal to the surface density, $K = 4\times 10^{-3}$ W $\mbox{m}^{-1}$ $\mbox{K}^{-1}$, $C=680$ J $\mbox{kg}^{-1} \mbox{K}^{-1}$, $\epsilon=0.88$ and $A=0.12$.

In the analysis of the simulation output we are interested in the particle orbits at the moment of escaping the co-orbital resonance. We therefore determine that a particle initially in libration about $\mbox{L}_{5}$ escapes when the quantity $\lambda - \lambda_{\rm Earth}$ changes sign while the escape criterion for horseshoe libration is that
\begin{equation}
\cos \left(\lambda - \lambda_{\rm Earth}\right) > 1 - \delta
\label{eq:esc}
\end{equation}
where $\delta \ll 1$ is a user-defined parameter. Some care should be exercised in choosing $\delta$ since too high a value may trigger false detections for the largest amplitude horseshoes, while too low a value would miss escape events due to the finite sampling resolution of the integration output. Through trial and error, we have chosen $\delta=2\times 10^{-3}$ as a good compromise between these competing requirements.

Particles may also transition between tadpole to horseshoe libration or from one triangular equilibrium point to the other. Transitions of the first type may not trigger the horseshoe escape criterion (Eq.~\ref{eq:esc}), however the onset of transitions in general signals an instability that promptly leads to escape \citep{Tsiganis.et.al2000,Connors.et.al2011,Dvorak.et.al2012} hence our escape detection procedure remains valid.

\section{\label{sec:results} Results}
\subsection{\label{sec:bxx} $N$-body runs}
\begin{figure}
\centering
\includegraphics[width=87mm,angle=0]{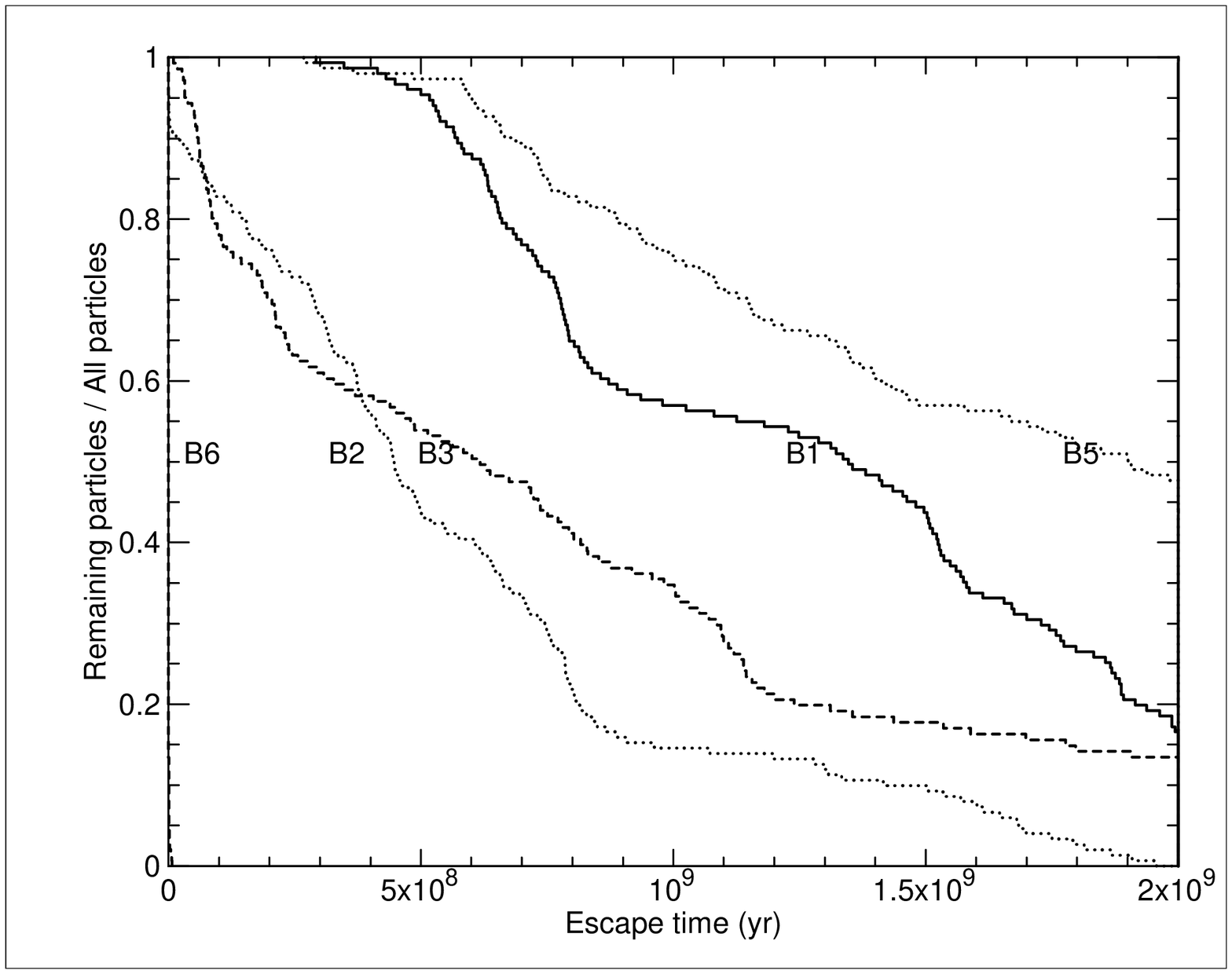}\\\includegraphics[width=87mm,angle=0]{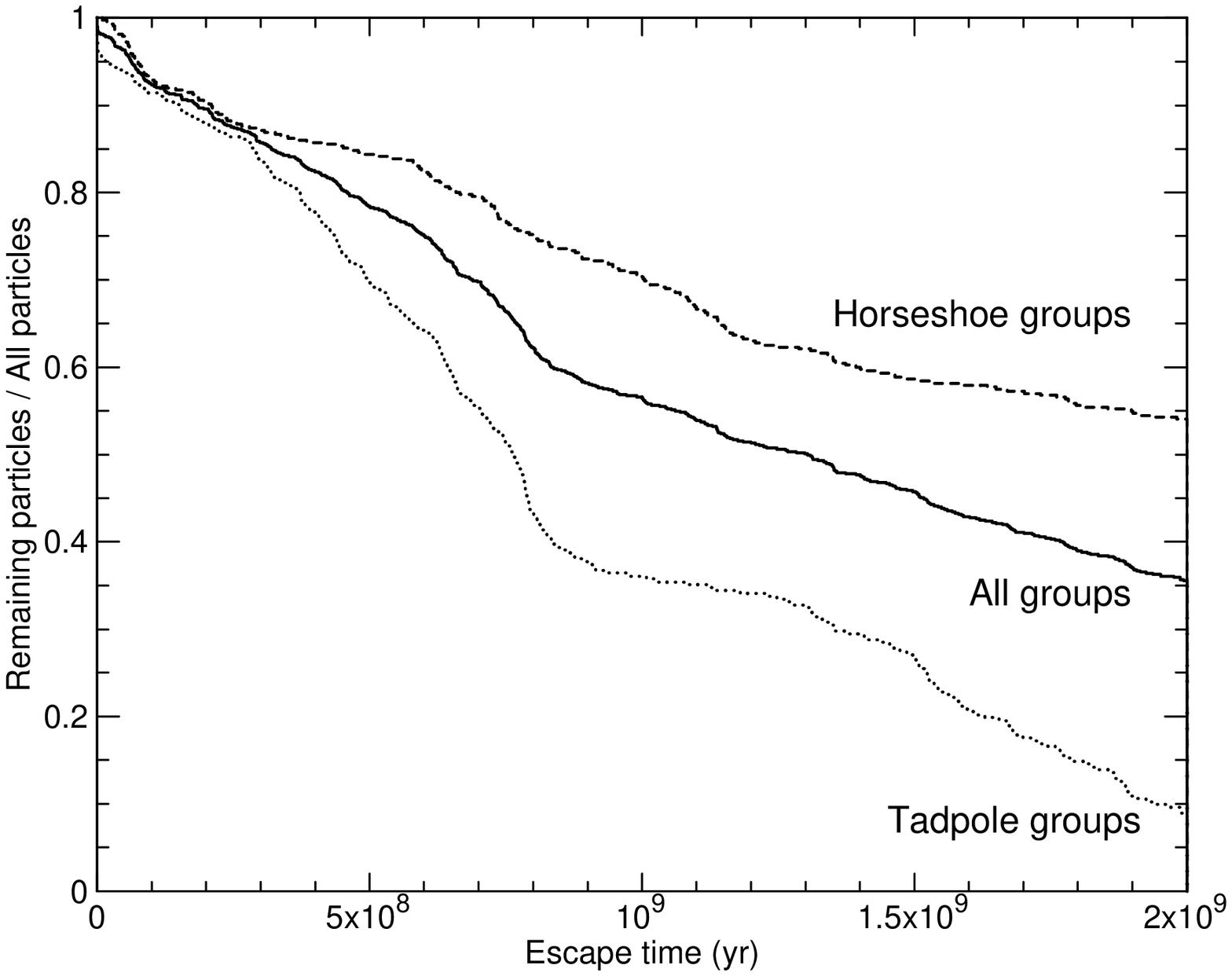}\\\includegraphics[width=87mm,angle=0]{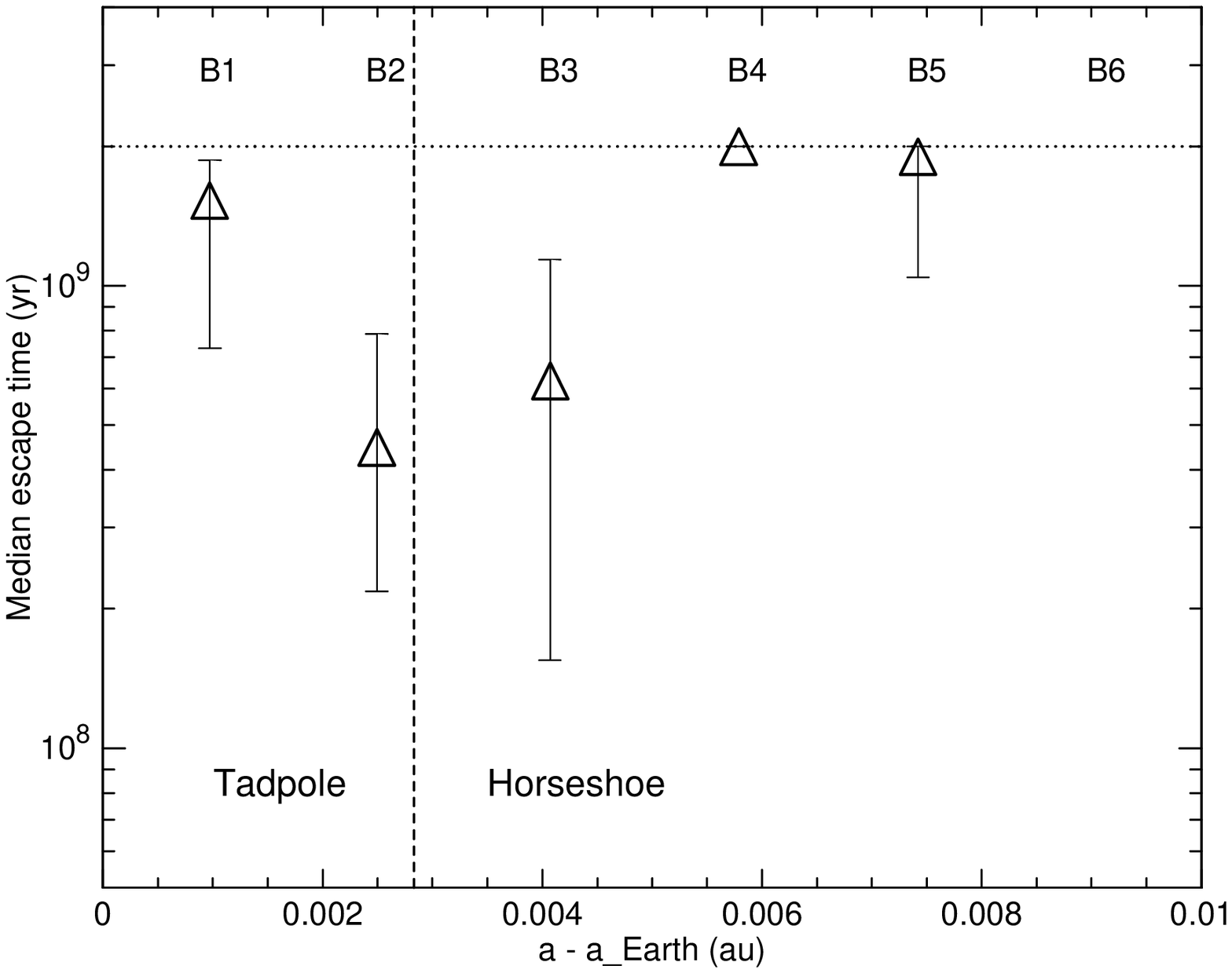}
\caption[{\bf Top}: Cumulative number of remaining coorbitals over time for the different coorbital test particle groups in the numerical simulations. All particles in group B6 escaped within the first $10^{7}$ yr of the simulations while all B4 particles remained in stable libration until the end of the runs. {\bf Middle}: As in top panel, but for the collective samples of tadpole, horseshoe and all co-orbital particles. {\bf Bottom}: Statistical estimates of the escape time for the same test particle groups as in the top panel. The triangle represents the median value, the error bars enclose the central 50\% of the sample and the dashed horizontal line indicates the duration of each run.]{{\bf Top}: Cumulative number of remaining coorbitals over time for the different coorbital test particle groups in the numerical simulations. All particles in group B6 escaped within the first $10^{7}$ yr of the simulations while all B4 particles remained in stable libration until the end of the runs. {\bf Middle}: As in top panel, but for the collective samples of tadpole, horseshoe and all co-orbital particles. {\bf Bottom}: Statistical estimates of the escape time for the same test particle groups as in the top panel. The triangle represents the median value, the error bars enclose the central 50\% of the sample and the dotted horizontal line indicates the duration of each run.}
\label{fig:escapes2}
\end{figure}

We show particle escape statistics from the B{[i]} simulations in Fig.~\ref{fig:escapes2}. The top panel shows the cumulative number of particles remaining in co-orbital libration as a function of time. The middle panel shows the same data as in the top panel but for the aggregated tadpole (groups B1 \& B2) and horseshoe (groups B3-B5; B6 particles readily escape, see below) particle sets. As the different simulation runs were completed, we found that the output files for some particles were corrupted in that they could not be read by the s/w included in MERCURY for this purpose. This prevented us from determining the dynamical evolution and final fate for 20 out of a total of 906 particles. The statistics we present here were derived from the remaining 886 particles. The fraction of particles ($\sim$2\%) affected by this issue is relatively tiny and unlikely to affect our conclusions. The bottom panel presents different statistical measures of the escape time for each group, namely the median, indicated by the triangle, and the central 50\% of the sample (error bars) obtained from the cumulative distributions. Here, the dashed vertical line represents the theoretical boundary between tadpole and horseshoe orbits (Tadpole-Horseshoe Boundary or THB) from Eq.~\ref{eq:thb}.

We observe considerable variation in the stability properties of different groups. The most extreme behaviour is observed for group B6 ($\Delta a$ = 0.0091 au), where all particles escape within the first $10^{7}$ yr of the simulation start, and group B4 ($\Delta a$ = 0.0058 au), where all particles remain in stable horseshoe libration for the full $2 \times 10^{9}$ yr. For the remaining groups, typical escape time - as quantified by the interquartile interval - is longer than a few times $10^{8}$ yr. At $t\simeq 8 \times 10^{8}$ yr, an elbow appears in the loss curves for groups B1 and B2; this is reflected in the overall tadpole loss curve in the middle panel. The rate of tadpole and horseshoe attrition is similar initially, however after a few times $10^{8}$ yr the distributions diverge with a faster loss rate for tadpoles. At the end of the simulations, 9\% (27/302) of $\mbox{L}_{5}$ tadpoles remain, compared to 54\% (236/435) for horseshoes (Table~\ref{tab:esc}) or a factor of six difference in loss efficiency. The overall 2 Gyr survival fraction of Earth coorbitals from our runs is then 28\%, assuming the same loss fraction for $\mbox{L}_{4}$ as found here for $\mbox{L}_{5}$ while the fraction of horseshoe vs tadpole survivors is $236/\left(2\times 27\right) \simeq 4.4$. A simple extrapolation of the former figure over the age of the solar system suggests a survival fraction of $\sim 8$\% for resident co-orbitals. \citet{Cuk.et.al2012} also found an increased survivability of horseshoe over tadpole particles but their investigation covered a shorter period, 700 Myr. We do not observe the asymptotic tails of long-lived coorbitals found in that work (cf Fig~3); the profiles in the middle panel of Fig.~\ref{fig:escapes2} are more consistent with either linear or piecewise-linear time dependence. The difference could be due to the different initial orbits for the particles and/or our coarser sampling of phase space compared to the \citeauthor{Cuk.et.al2012} study.

The fast instability observed for group B6 shows that the effective 2 Gyr stability threshold for Earth horseshoes must lie somewhere between 0.0074 and 0.0091 au. This is in agreement with \citet{Cuk.et.al2012} who found that horseshoe libration at $\Delta a$ $\simeq$0.0075 au persists for 700 Myr and further suggests that the domain of stable horseshoe libration found in that work is not significantly eroded for 3$\times$ the simulation time and $\sim$50\% the age of the solar system.

In examining the orbit evolution, short-term variation of the osculating orbits would mask the slow orbit diffusion we want to study. We filter out these rapid variations by applying a boxcar average to the numerical output with boxcar widths of $5\times 10^5$ yr for the eccentricity and $3\times 10^{5}$ yr for the inclination. At the same time, we monitor the change in the amplitude of the semimajor axis libration around 1 au by recording the difference between the minimum and maximum value with a time resolution of $6\times10^{4}$ yr.

\begin{figure}
\hspace{-5mm}\includegraphics[width=87mm,angle=0]{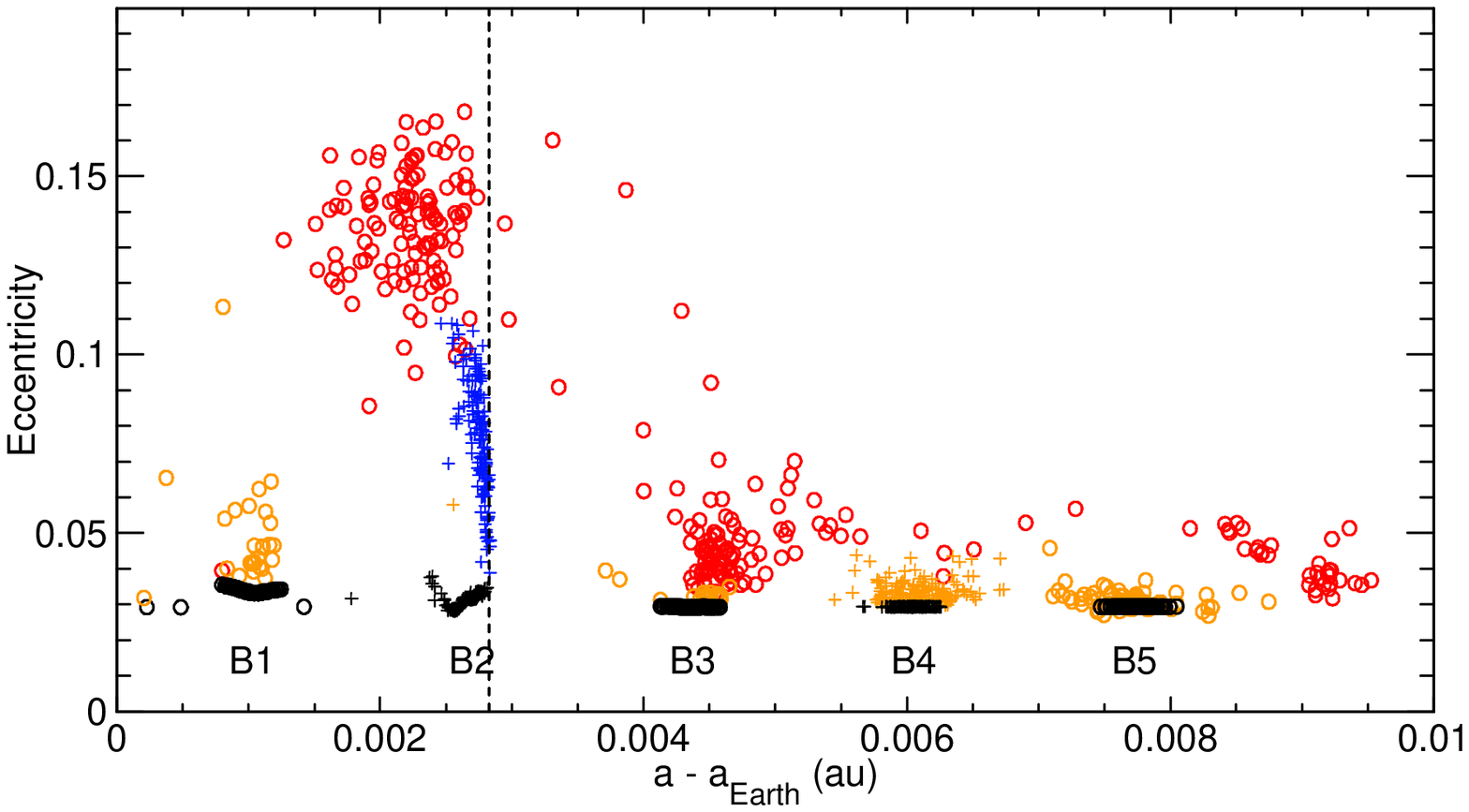}
\includegraphics[width=87mm,angle=0]{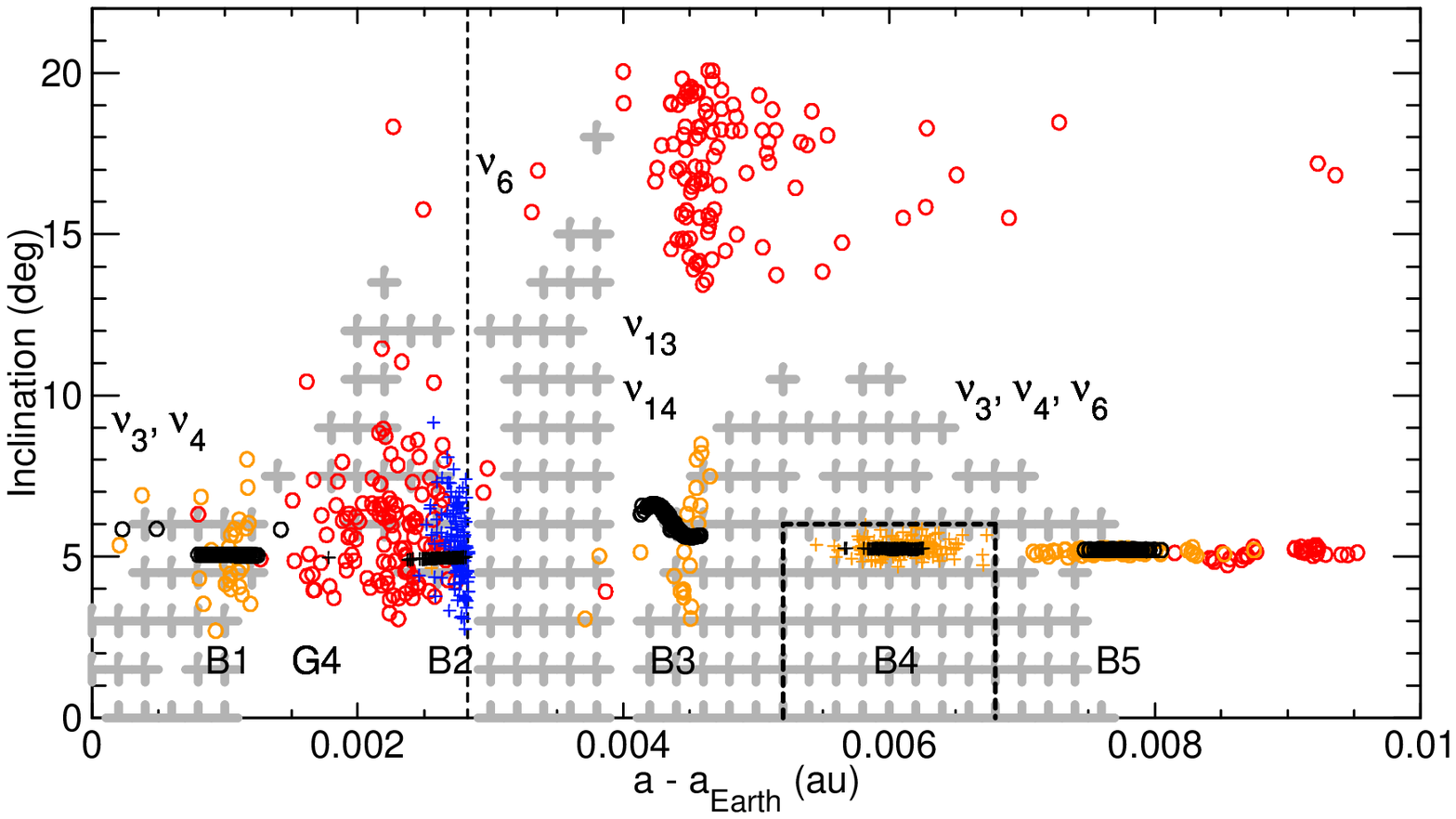}
\caption[Initial and final states of particles in the $N$-body runs. Black points represent the average orbits in the first $5\times 10^{5}$ yr, all other points represent final states. A red colour indicates the location of an escaping particle at the moment libration of the critical argument is broken, an amber colour shows surviving particles at the end of the run. To distinguish between particles in groups B1 and B2, we use a blue colour to indicate escaping particles from B2. The dashed vertical line indicates the theoretical Tadpole-Horseshoe Boundary (THB) in the planar, circular case. The figure is annotated with the locations of secular resonances from \citet{Zhou.et.al2019}; the gray fenced region represents locations of surviving co-orbital particles in the 700 Myr simulations of \citet{Cuk.et.al2012}. The rectangle centered at $\Delta a$=0.006 au demarcates a region where resident planetesimals may survive over the age of the solar system.]{Initial and final states of particles in the $N$-body runs. Black points represent the average orbits in the first $5\times 10^{5}$ yr, all other points represent final states. A red colour indicates the location of an escaping particle at the moment libration of the critical argument is broken, an amber colour shows surviving particles at the end of the run. To distinguish between particles in groups B1 and B2, we use a blue colour to indicate escaping particles from B2. The dashed vertical line indicates the theoretical Tadpole-Horseshoe Boundary (THB) in the planar, circular case. The figure is annotated with the locations of secular resonances from \citet{Zhou.et.al2019}; the gray fenced region represents locations of surviving co-orbital particles in the 700 Myr simulations of \citet{Cuk.et.al2012}. The rectangle centered at $\Delta a$=0.006 au demarcates a region where resident planetesimals may survive over the age of the solar system.}
\label{fig:l_vs_e_vs_i}
\end{figure}

Figure~\ref{fig:l_vs_e_vs_i} shows the orbital locations of test particles at different times in the simulations. Black symbols mark the initial locations, red or blue points indicate the particle locations at the moment the co-orbital resonance is broken and amber points show the final orbit of surviving particles. We observe that orbits diffuse by different amounts and in different directions depending on the starting location. A qualitative difference between the eccentricity and inclination diffusion is that inclination diffuses to both higher and lower values while the eccentricity diffuses to higher values only.

To help place our results in context, we superimpose in the bottom panel of this figure the loci of surviving co-orbital particles in the 700 Myr simulations of \citet{Cuk.et.al2012} as a gray fenced pattern. We also indicate the approximate locations of secular resonances from \citet{Zhou.et.al2019} with the same notation used in that work: ``$\nu_{x}$'' and ``$\nu_{1x}$'' for linear eccentricity- and inclination-type resonances respectively and ``$G_{x}$'' for higher order eccentricity-type resonances. With the exception of group B4, the initial locations of particles in our runs generally straddle boundaries between stable and unstable motion. A moderate fraction (14-48\%) of particles in each group escapes during the simulations, except for group B4 where no escapes are observed and for group B2 where all but one particle escape. This further reinforces the conclusion that the domain mapped out by \citeauthor{Cuk.et.al2012} is not significantly eroded over longer time periods than those explored in their numerical runs.

For tadpole particles (groups B1 and B2) we find that escape occurs as orbits diffuse to higher values of $e$ and $\Delta a$ and that no particles in these groups diffuse beyond the THB. For group B1, particles typically escape when $e$=0.1-0.17 (red points) while group B2 escapees lie along the THB with $e$=0.04-0.1 (blue points). Therefore, proximity of the initial states to the THB is important in determining how tadpole asteroids leave the stable domain, while eccentricity excitation beyond $e \simeq 0.1$ caused loss of Trojans with vanishing initial $\Delta a$. We note that such orbits are still not eccentric enough to allow physical approaches to the Earth or other planets, therefore we suggest that the instability is actually caused by the eccentricity-type secular resonances abundant in the tadpole region \citep{Zhou.et.al2019}.

The dynamical evolution of horseshoe particles shows a different character as they do not generally reach the high-eccentricity states of escaping Trojans. Group B3 lies adjacent to the THB and to the left of the stable region identified in \citet{Cuk.et.al2012}. Here, all escaping particles have inclinations $I$=$15^{\circ}$-$20^{\circ}$. In analogy, therefore, to the escape of group B1 Trojans facilitated by the eccentricity-type resonances, here the mechanism that breaks horseshoe libration appears linked to the two inclination-type secular resonances present at $\Delta a \sim 0.004$ au. To the right of the stable region, B5 particles represent the largest amplitude horseshoes in our simulations and show strong diffusion in $\Delta a$. All escaping particles have higher $\Delta a$ than the initial states and $\gtrsim 0.008$ au, in agreement with past studies of the stability threshold for Earth horseshoes \citep{WeissmanWetherill1974,Cuk.et.al2012,Zhou.et.al2019}. Still, about half of all particles in the group survive as horseshoes for 2 Gyr, one such particle finishes the simulation at $\Delta a \simeq 0.0087$ au.


Group B4 lies deep within a stable island between 0.004 and 0.0075 au. These particles experience weak to moderate orbital diffusion and all remain in the resonance for the entire duration of the simulations. We can use this result to constrain the initial number of asteroids residing in the stable domain. Strictly speaking,  these constraints apply only to asteroids large enough to be unaffected by size-dependent forces ($\gg 1$ km, Section~\ref{sec:by}) and we refer to these objects as ``planetesimals'' to distinguish them from the smaller asteroids.

\begin{figure}
\centering
\includegraphics[width=87mm,angle=0]{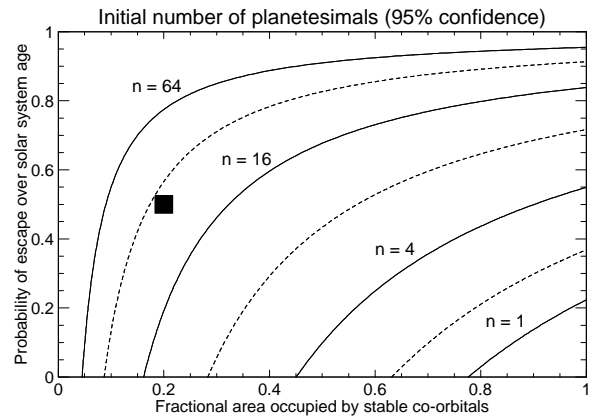}
\caption[Initial number $n$ of Earth co-orbital planetesimals given that none are presently observed from Eq.~\ref{eq:n0_given_n_eq_0} and a confidence level of $0.95$ as a function of the fractional area $r$ of the phase space region where such planetesimals survive for the age of the solar system with probability $1 - q$. Planetesimals are certain to escape from the remaining phase space. The filled square corresponds to the case with $r=0.2$ and $q=0.5$ discussed in the text.]{Initial number $n$ of Earth co-orbital planetesimals given that none are presently observed from Eq.~\ref{eq:n0_given_n_eq_0} and a confidence level of $0.95$ as a function of the fractional area $r$ of the phase space region where such planetesimals survive for the age of the solar system with probability $1 - q$. Planetesimals are certain to escape from the remaining phase space. The filled square corresponds to the case with $r=0.2$ and $q=0.5$ discussed in the text.}
\label{fig:cdf_r_vs_q}
\end{figure}

First, we assume that planetesimals within a rectangular region (Fig.~\ref{fig:l_vs_e_vs_i}) with $I \leq 6^{\circ}$ and $0.0052 \leq \Delta a \leq 0.0068$ au share the same stability properties as the B4 particles and $(1 -q)$$\times$$100$\% odds that these particles will survive over the age of the solar system. We then divide the entire stable domain mapped out by \citet{Cuk.et.al2012} in two separate regions, one outside the rectangle where objects are certain to escape over the age of the solar system ie $q=1$ and the other within the rectangle where the escape probability is $0<q<1$. The area of the latter region normalised by the entire domain mapped out by \citeauthor{Cuk.et.al2012} is represented by the parameter $r$. The probability that at most $n$ objects were present at $t=0$ given that none are observed at $t=4.5$ Gyr is then
\begin{equation}
\label{eq:n0_given_n_eq_0}
P(N(0) \leq n\mbox{ }|\mbox{ }N(4.5\mbox{ Gyr}) = 0) = 1 - (1 - r + r q)^{n+1}
\end{equation}
where the extra unit added to the exponent accounts for the event $N(0) = 0$ ie no Trojans were initially present. Level curves of Eq.~\ref{eq:n0_given_n_eq_0} 
correspond to statistical confidence levels for the initial number $n$ of co-orbital planetesimals as functions of $r$ and $q$. In Fig.~\ref{fig:cdf_r_vs_q} we show these constraints for 95\% confidence. A vanishing initial number of planetesimals ($n\leq 1$, bottom right area of the plot) presupposes that most of the phase space is stable ie $r\simeq 1$ and that the probability of escape from the remaining phase space is low. On the other hand, hundreds of planetesimals can be allowed for (top left area of plot) if the extent of the $r$-region is limited ($r\sim 0$) {\it or} if planetesimals readily escape from the remaining phase space ie $q\sim 1$. 

The chosen rectangular region in Fig.~\ref{fig:l_vs_e_vs_i} corresponds to 20\% of the stable domain of \citeauthor{Cuk.et.al2012}, where the area left of the THB was counted twice to allow for the number of equilibrium points. For $r=0.2$ and by further assuming $q=0.5$ we obtain $n=27$ at 95\% confidence, this is represented by the filled square in Fig.~\ref{fig:cdf_r_vs_q}. To assess the local parameter sensitivity of the constraints, we vary $r$ by 50\% of the reference value, that is $r=0.2^{+0.1}_{-0.1}$ while keeping $q$ fixed to obtain $n=27^{-9}_{+30}$ at 95\% and $n=42^{-15}_{+46}$ at 99\% confidence respectively. The median for the same choices of parameter values is $n=5^{-2}_{+7}$. We obtain the same numerical bounds if we fix $r$ at $0.2$ and instead vary $q$ over the range $0.25-0.75$, therefore our determination of $n$ appears locally insensitive to the parameters.

\begin{figure}
\centering
\includegraphics[width=87mm,angle=0]{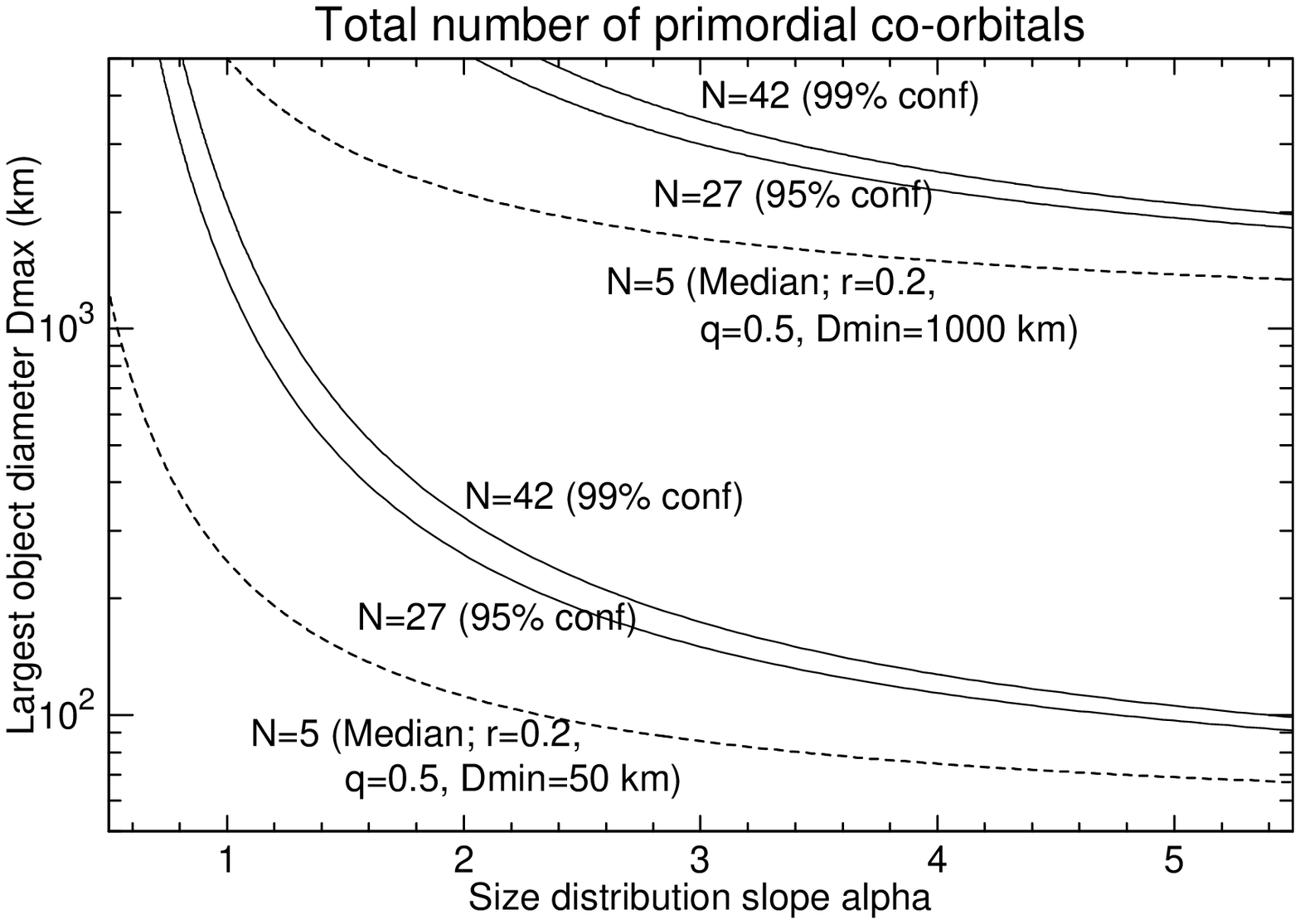}\\\includegraphics[width=87mm,angle=0]{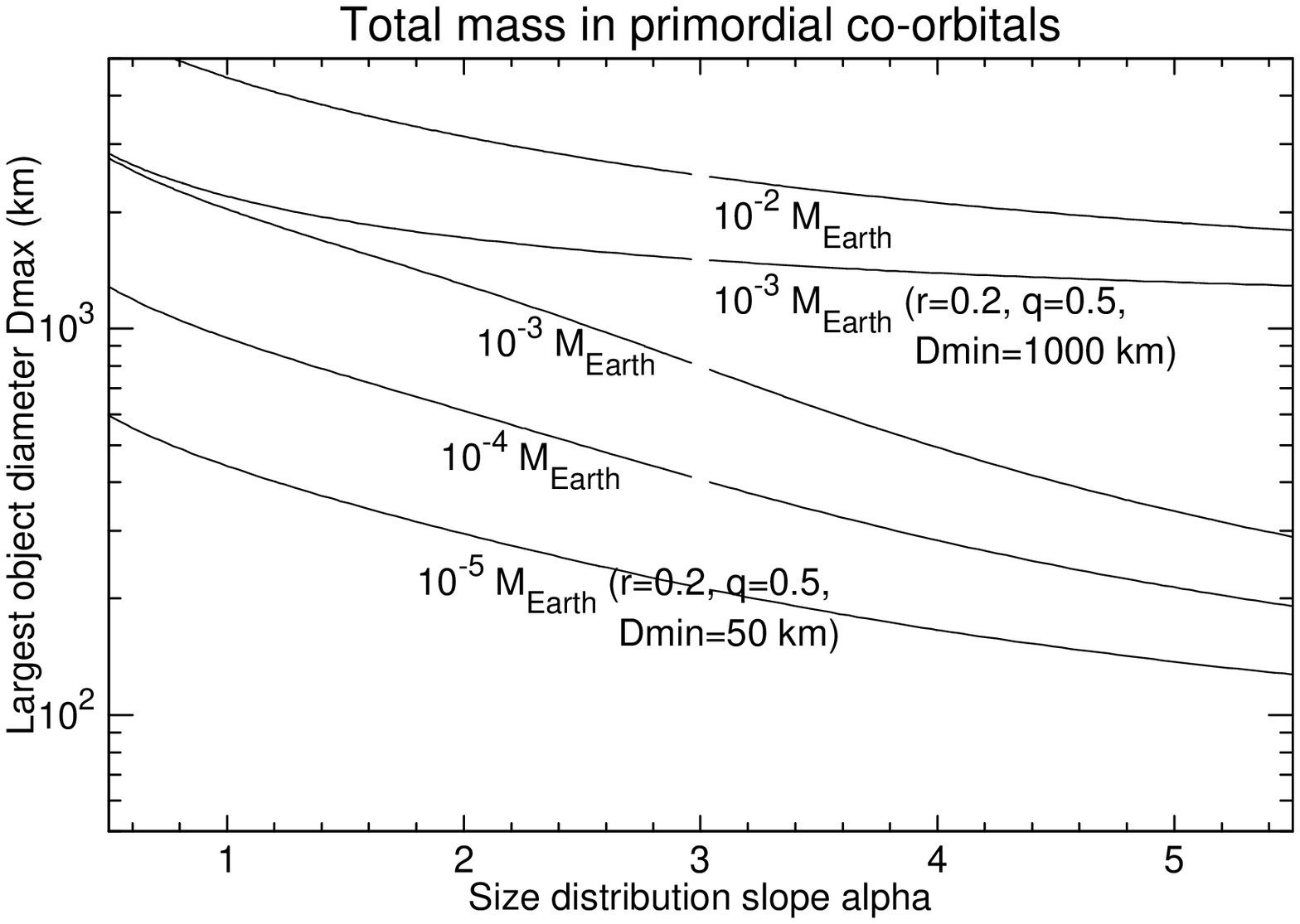}
\caption[Population properties of primordial co-orbitals from statistical analysis of particle escapes in our numerical simulations. {\bf Top}: Constraints on total number of co-orbitals as a function of the assumed size distribution slope $\alpha$ and largest object diameter $D_{\rm max}$. Two different populations are shown, one where $D_{\rm min}=50$ km and $D_{\rm max}\leq 1000$ km and another where $D_{\rm min}=1000$ km and $D_{\rm max}\leq5000$ km. {\bf Bottom}: Total mass in units of Earth masses for the two populations shown in the top panel.]{Population properties of primordial co-orbitals from statistical analysis of particle escapes in our numerical simulations. {\bf Top}: Constraints on total number of co-orbitals as a function of the assumed size distribution slope $\alpha$ and largest object diameter $D_{\rm max}$. Two different populations are shown, one where $D_{\rm min}=50$ km and $D_{\rm max}\leq 1000$ km and another where $D_{\rm min}=1000$ km and $D_{\rm max}\leq5000$ km. {\bf Bottom}: Total mass in units of Earth masses for the two populations shown in the top panel.}
\label{fig:alpha_vs_dmax}
\end{figure}

These constraints can be interpreted in terms of the likely population properties of the co-orbitals. If the number of asteroids larger than a given size $N(>D)$ follows a power law with slope $\alpha$ then the total number $N_{\rm tot}$ and mass $M_{\rm tot}$ of co-orbitals are given by
\begin{eqnarray}
N_{\rm tot} &=& {\left(\frac{D_{\rm min}}{D_{\rm max}}\right)}^{-\alpha} \label{eq:number_mass_vs_d_1}\\ M_{\rm tot} &=& \frac{\alpha}{6(\alpha-3)} \pi \rho D^{3}_{\rm max} \left( \left(\frac{D_{\rm min}}{D_{\rm max}}\right)^{3-\alpha} - 1 \right)
\label{eq:number_mass_vs_d_2}
\end{eqnarray}
where $D_{\rm min}$ and $D_{\rm max}$ are the respective sizes of the smallest and largest object and $\rho$ is the bulk density. For the degenerate case $\alpha=3$ the expression for $M_{\rm tot}$ assumes a different form. In Fig~\ref{fig:alpha_vs_dmax}, top panel we show these number constraints for the nominal case $r=0.2$ and $q=0.5$ as functions of $D_{\rm max}$ and $\alpha$. We consider two population types, one where $D_{\rm min} = 50$ km and another where $D_{\rm min} = 1000$ km for $D_{\rm max}$ up to 5000 km in either case. A slope $\alpha=2.5$ represents a population in collisional equilibrium \citep{Dohnanyi1969,Bottke.et.al2015}. Slope values of $\sim$3.5 are appropriate for the largest Main Belt asteroids \citep{Bottke.et.al2005} while shallow, sub-collisional slopes may apply to planetesimals that participated to an early impact bombardment of the terrestrial planets \citep{Bottke.et.al2007b,Bottke.et.al2010}. A limitation of this approach is that it cannot be used to constrain initial co-orbital populations with abundant small ($D_{\rm min} = 1$ km or smaller, see Section~\ref{sec:by}) asteroids for which size-dependent forces, rather than gravitational diffusion, should be the dominant loss mechanism. In particular, such objects would dominate  the mass budget for steep ($\alpha > 3$) size distributions.

We find that object sizes much larger than few$\times$$D_{\rm min}$ are generally not likely under our constraints, except perhaps for flat ($\alpha \lesssim 2$) size distributions with a low ratio of small vs large objects. This is probably the result of imposing the constraint of a low number of objects for which a size distribution is not well-defined. The strongest statement we can therefore make is that the concept of a co-orbital population composed of similar-sized objects is a reasonable approximation to the truth, independently of the shape of the distribution.

Our constraints cannot discriminate between populations dominated by either large (1000-km class) or small (100-km class) objects, yet these would translate into very different mass budgets (Fig~\ref{fig:alpha_vs_dmax}, bottom panel). If we use the median contour from the top panel as a guide, we see that masses of $< 10^{-5}$ $M_{\rm Earth}$ are implied for a population of 100-km-sized objects and $< 10^{-2}$ $M_{\rm Earth}$, ie less than a lunar mass, for a population of $1000$-km-sized objects, where we have assumed $\rho = 2700$ kg $\mbox{m}^{-3}$. Both estimates can be accommodated within the expected planetesimal mass at 1 au at the time of the formation of the Earth-Moon system \citep{Weidenschilling1977,Bottke.et.al2007b}.

\subsection{\label{sec:dearth}Coupling to the Earth's orbit}
\begin{figure}
\centering
\includegraphics[width=87mm,angle=0]{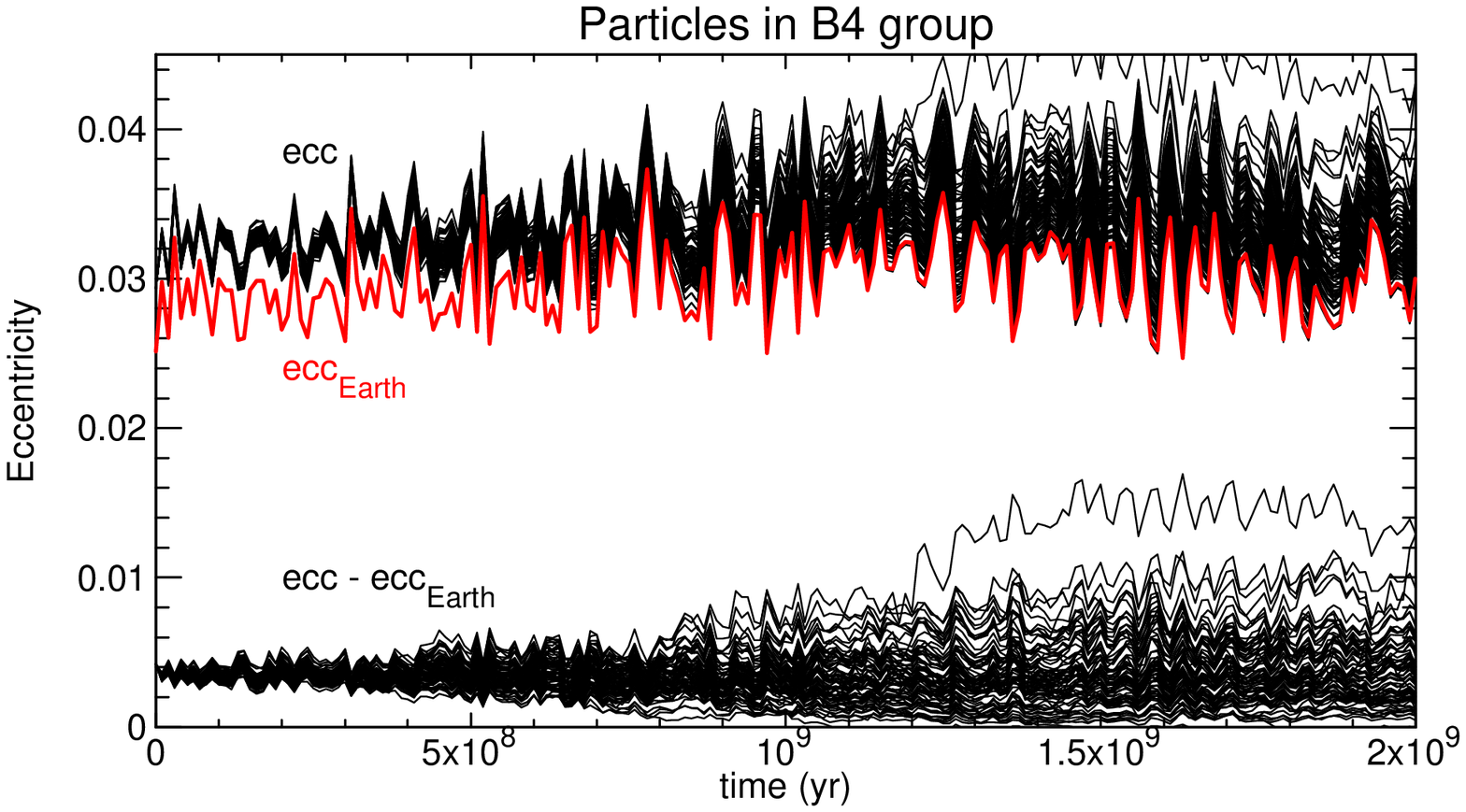}\\\includegraphics[width=87mm,angle=0]{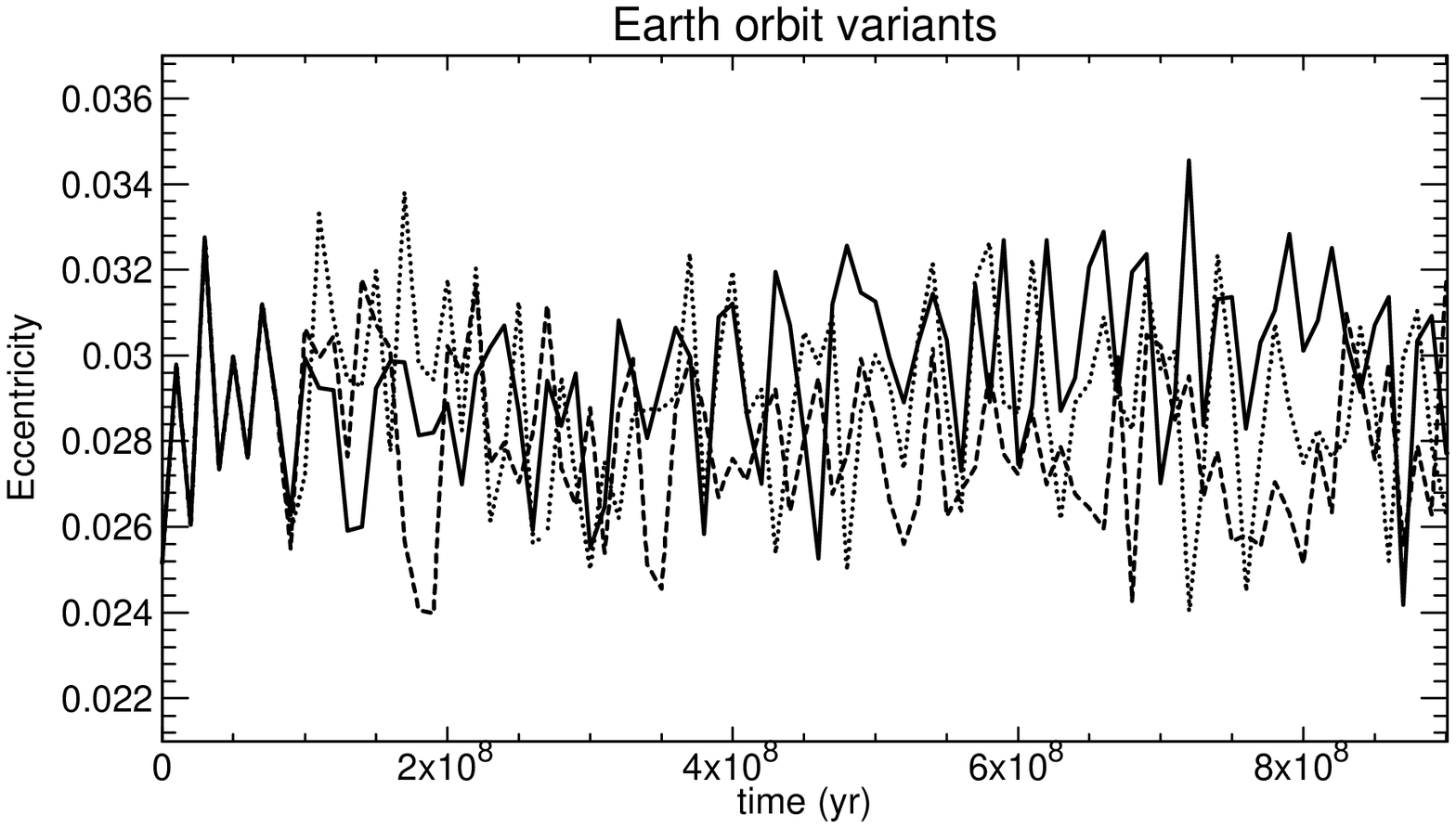}
\caption[{\bf Top}: Eccentricity evolution for particles in group B4. The red curve shows the concurrent evolution of Earth's eccentricity during the simulations. {\bf Bottom}: Eccentricity evolution for the three Earth orbit variants used in the B4E simulation.]{{\bf Top}: Eccentricity evolution for particles in group B4. The red curve shows the concurrent evolution of Earth's eccentricity during the simulations. {\bf Bottom}: Eccentricity evolution for the three Earth orbit variants used in the B4E simulation.}
\label{fig:ecc_earth}
\end{figure}

On the longest timescales, the osculating eccentricity is formed by superposition of an intrinsic (or free) component on a forced term \citep{MurrayDermott1999}.  This forced eccentricity is approximately the same for the Earth and the coorbitals \citep{Morais1999,Georgakarakos.et.al2016} while the free eccentricity diffuses to either higher or lower values over time. To isolate the long-term variations we are interested in, the numerical output has been boxcar-averaged with a 10 Myr window and we make use of this running average, instead of the osculating value, for the remainder of this section. Detecting escapes from the co-orbital resonance is, however, still done through the osculating output and Eq.~\ref{eq:esc}. We demonstrate the dynamics in the top panel of Fig~\ref{fig:ecc_earth} by showing the eccentricity of the particles, that of the Earth as well as their difference. The asteroid orbits are therefore coupled to the Earth's own eccentricity history and will always have $e > e_{\rm Earth}$. 

These observations suggest that the survival of primordial co-orbitals to the present day must depend to some degree on the Earth's orbital history. In our simulations we observe that the Earth's average eccentricity varies between 0.025 to 0.037, a relatively narrow range. Given that the orbit evolution of our planet is stochastic over timescales of Gyr \citep{Laskar1994}, actual past orbit variations in excess of those noted here cannot be ruled out. In particular, episodes when $e_{\rm Earth}$ is significantly greater than $0.04$ over the past 4 Gyr, if they occurred, may have triggered increases in the rate of depletion of asteroids from the Trojan reservoirs.

To investigate further, took the first 51 entries in the list of 151 particle initial conditions generated for the B1 and B4 group simulations, modified the y-component of the Earth's initial cartesian position vector by multiples of $10^{-9}$ au to generate three distinct Earth orbit variants per group and re-ran the simulations for those particles until $t=900$ Myr, referring to these runs as B1E (tadpoles) and B4E (horseshoes). The eccentricity evolution of the different Earth orbit variants in the B4E runs is shown in the bottom panel of Fig.~\ref{fig:ecc_earth}. None of the horseshoe coorbitals escaped while the number of escaping tadpole co-orbitals were 7, 15 \& 5. In the corresponding 2 Gyr simulations (ie B1 and B4), the number of escaping Trojan particles at $t=900$ Myr was 62, therefore the equivalent number of escapes if we were to begin that simulation with 51 instead of 151 particles would be [62$\times$(51/151)] = 20, somewhat higher but still compatible with the new runs. At the same time, the statistical properties of the time series for the different Earth orbit variants are essentially identical, in particular the maximum value of $e_{\rm Earth}$ among the three runs varied between 0.033 and 0.036. We conclude therefore that, for the current solar system architecture, excursions in the Earth's orbit eccentricity large enough to produce bulk destabilisation of deep co-orbitals, must not be very likely.

\subsection{\label{sec:by} Evolution under the Yarkovsky effect}
To find out how the Yarkovsky effect modifies the stability properties of long-lived Earth co-orbitals, we have re-integrated groups B1 and B4 with the Yarkovsky acceleration switched on. To distinguish between these simulations and the gravity-only runs in Section~\ref{sec:bxx} we refer to them as B1Y and B4Y respectively. As stated in Section~2, the magnitude of the Yarkovsky acceleration in these runs varied from $-17$ to $+17 \times 10^{-3}$ au $\mbox{Myr}^{-1}$ in terms of the equivalent drift in semimajor axis $a$, corresponding to a minimum object size of $D=50$ m for suitable parameter choices in the force model.

Switching on the Yarkovsky effect generally accelerates the loss of both tadpole and horseshoe co-orbitals. The median lifetime of B1Y particles is 3.6$\times$$10^{8}$ yr with all particles escaping before the end of the run. By comparison, the median lifetime in the B1 group was 1.5$\times$$10^{9}$ yr with 26/151 particles remaining in resonance at the end of the simulation. For the B4Y group of horseshoe particles, all but three particles escaped with a median escape time of 4.2$\times$ $10^{8}$ yr, similar to the B1Y runs. Our simulations therefore suggest that Earth coorbital asteroids of sizes of tens to hundreds of m are cleared over 2 Gyr with an efficiency that approaches 100\% regardless of the libration type.

\begin{figure}
\hspace{-5mm}\includegraphics[width=87mm,angle=0]{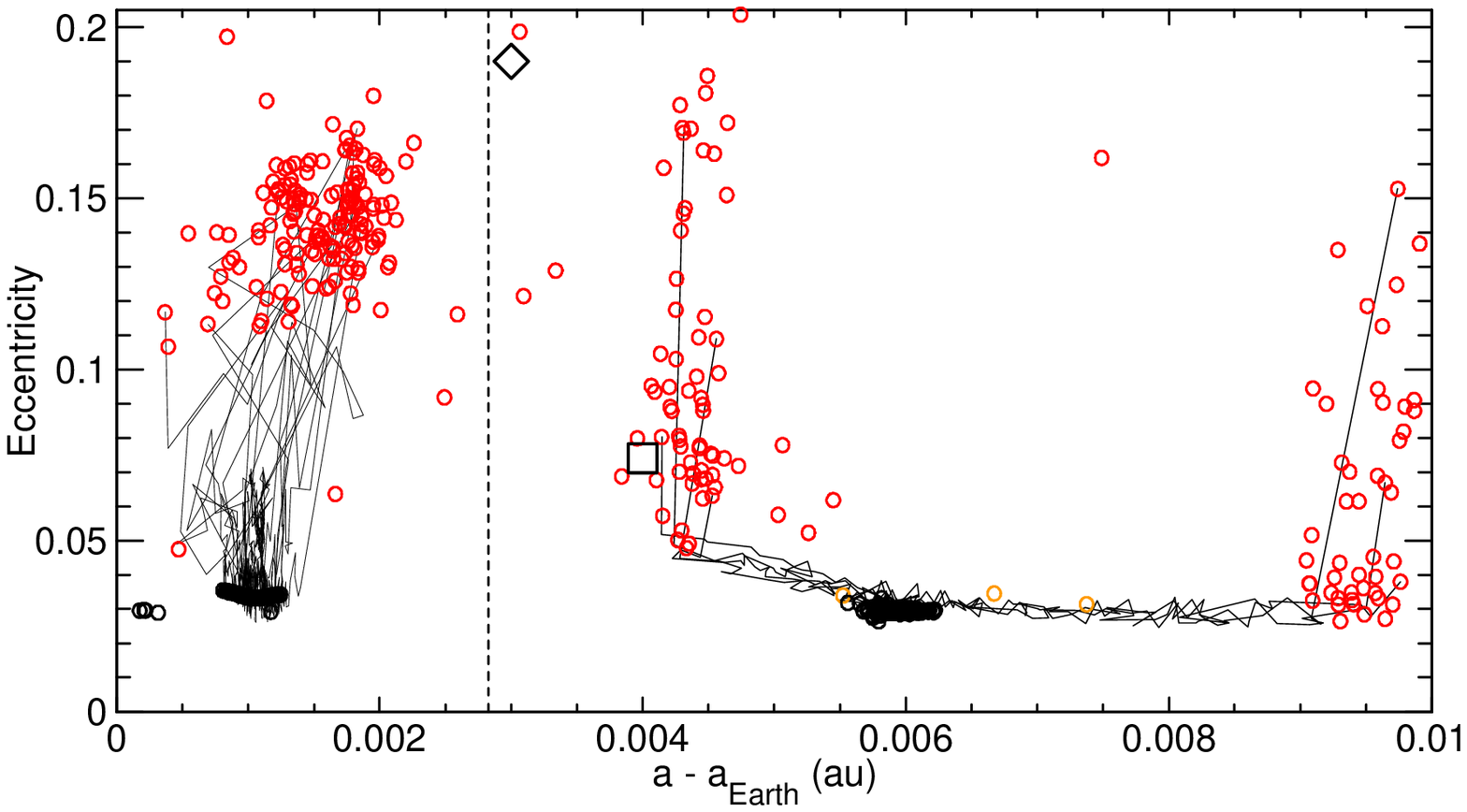}
\includegraphics[width=87mm,angle=0]{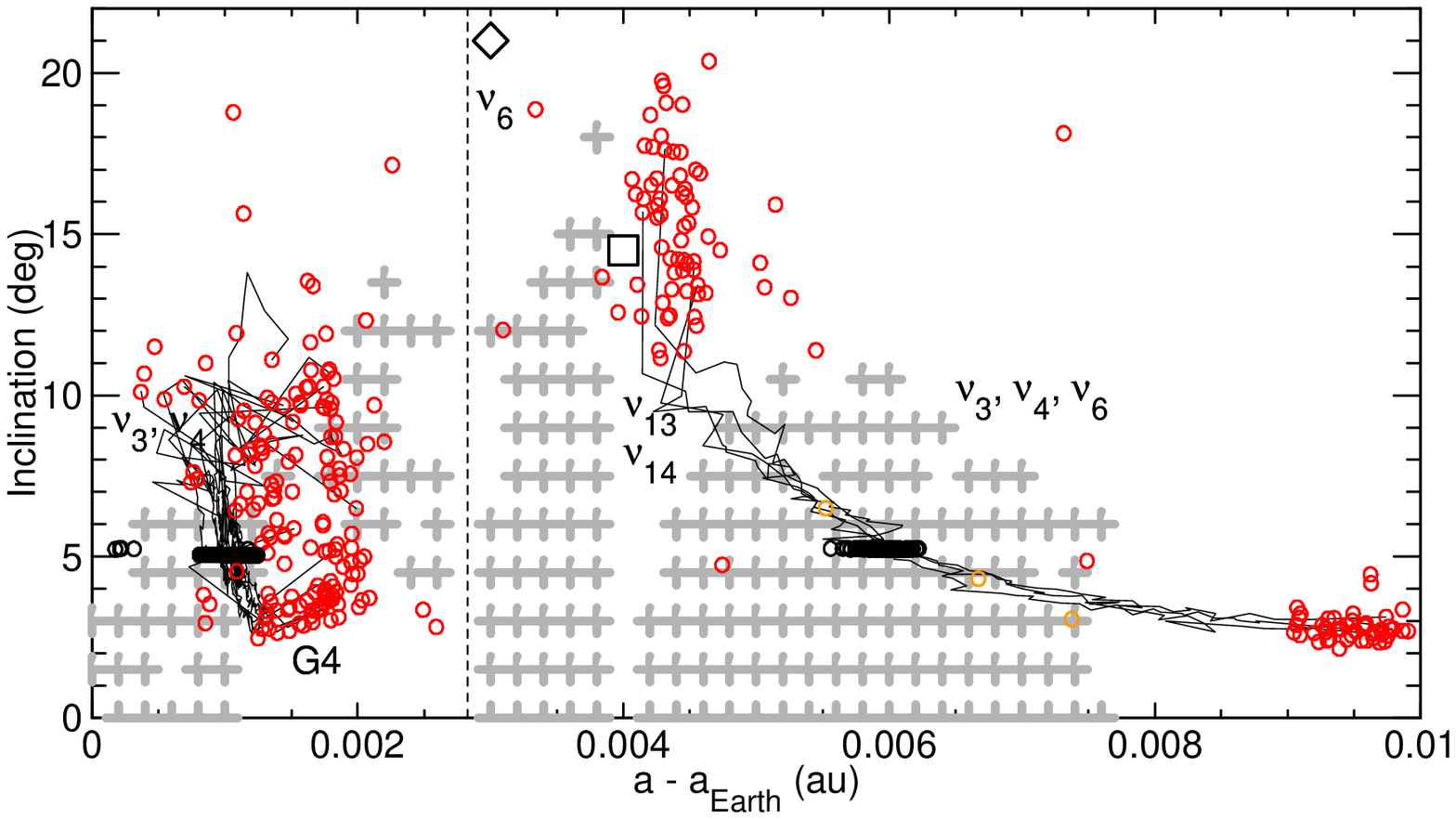}
\caption[Initial and end states of tadpole and horseshoe particles in the Yarkovsky simulations. Notation is as in Fig.~\ref{fig:l_vs_e_vs_i}. Black lines show the time evolution of the orbit for a selection of particles in each group. The diamond and open square symbols indicate the current orbits of asteroids 2010 $\mbox{TK}_{7}$ and (419624) 2010 $\mbox{SO}_{16}$ respectively.]{Initial and end states of tadpole and horseshoe particles in the Yarkovsky simulations. Notation is as in Fig.~\ref{fig:l_vs_e_vs_i}. Black lines show the time evolution of the orbit for a selection of particles in each group. The diamond and open square symbols indicate the current orbits of asteroids 2010 $\mbox{TK}_{7}$ and (419624) 2010 $\mbox{SO}_{16}$ respectively.}
\label{fig:l_vs_e_vs_i_yark}
\end{figure}

As with the gravitational simulations, we find that particles escape when they reach the boundary of the stability domain, however the route taken to reach that boundary is different (Fig~\ref{fig:l_vs_e_vs_i_yark}) with particles from each group following a common, deterministic path in $\Delta a$-$e$-$I$ space. This is more clearly seen for the horseshoe particles which we highlight by plotting a selection of trajectories. For particles with positive Yarkovsky acceleration, the libration amplitude $\Delta a$ decreases while $e$ \& $I$ increase; a negative acceleration has the opposite effect. The correlation between the sign of the change in the orbit elements and the direction of the Yarkovsky acceleration is similar to that observed in \citet{Cuk.et.al2015} for Mars Trojans and likely a generic property of Yarkovsky-driven orbit evolution in the co-orbital resonance \citep{WangHou2017}. The evolution of tadpole particles is qualitatively the same with horseshoes, though with higher overall change in $e$ and a smaller change in $a$. As the eccentricity of horseshoe particles is rapidly excited once they leave the stable region, the same process should be responsible for much of the observed eccentricity change of tadpole particles also.

In the same figure we show the respective locations of asteroids 2010 $\mbox{TK}_{7}$ and 2010 $\mbox{SO}_{16}$ as the open diamond and square symbols respectively. These asteroids are temporarily trapped in the co-orbital resonance and we see that they both occupy the boundaries of the stable regions, their residence probably facilitated by the existence of the secular resonances. In particular, the present orbit of 2010 $\mbox{SO}_{16}$ is similar to those particles escaping from group B4 and, in this sense, it is possible that this asteroid began as a deep horseshoe co-orbital that reached its present unstable orbit through gravitational diffusion assisted by the Yarkovsky effect. However, in the absence of compelling evidence for an extant population of deep co-orbitals to act as a source of such objects, the origins of 2010 $\mbox{SO}_{16}$ and $\mbox{TK}_{7}$ are probably more prosaic, having instead arrived from one or more of the principal NEO source regions \citep{MoraisMorbidelli2002}.

\begin{figure}
\centering
\includegraphics[width=87mm,angle=0]{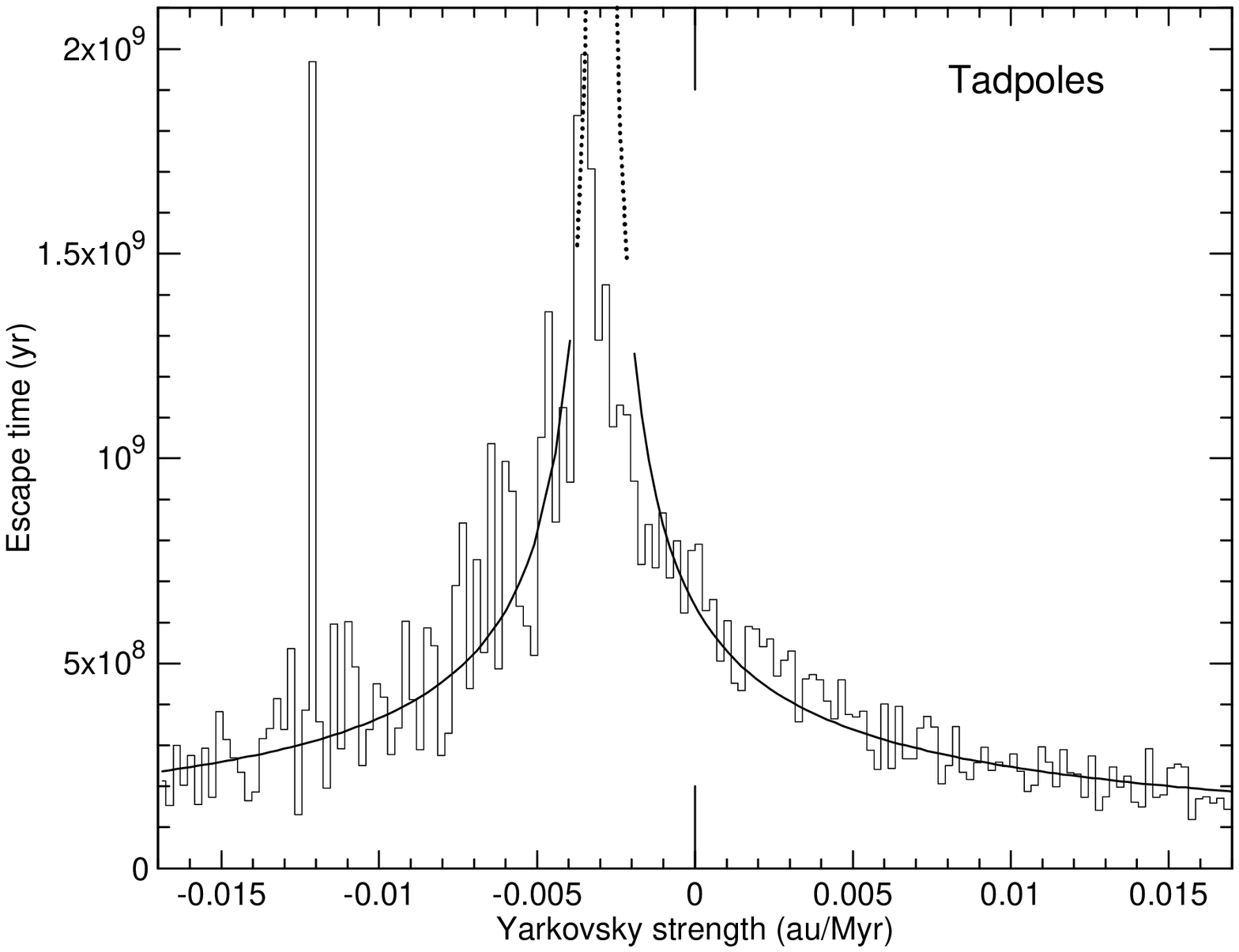}\\
\includegraphics[width=87mm,angle=0]{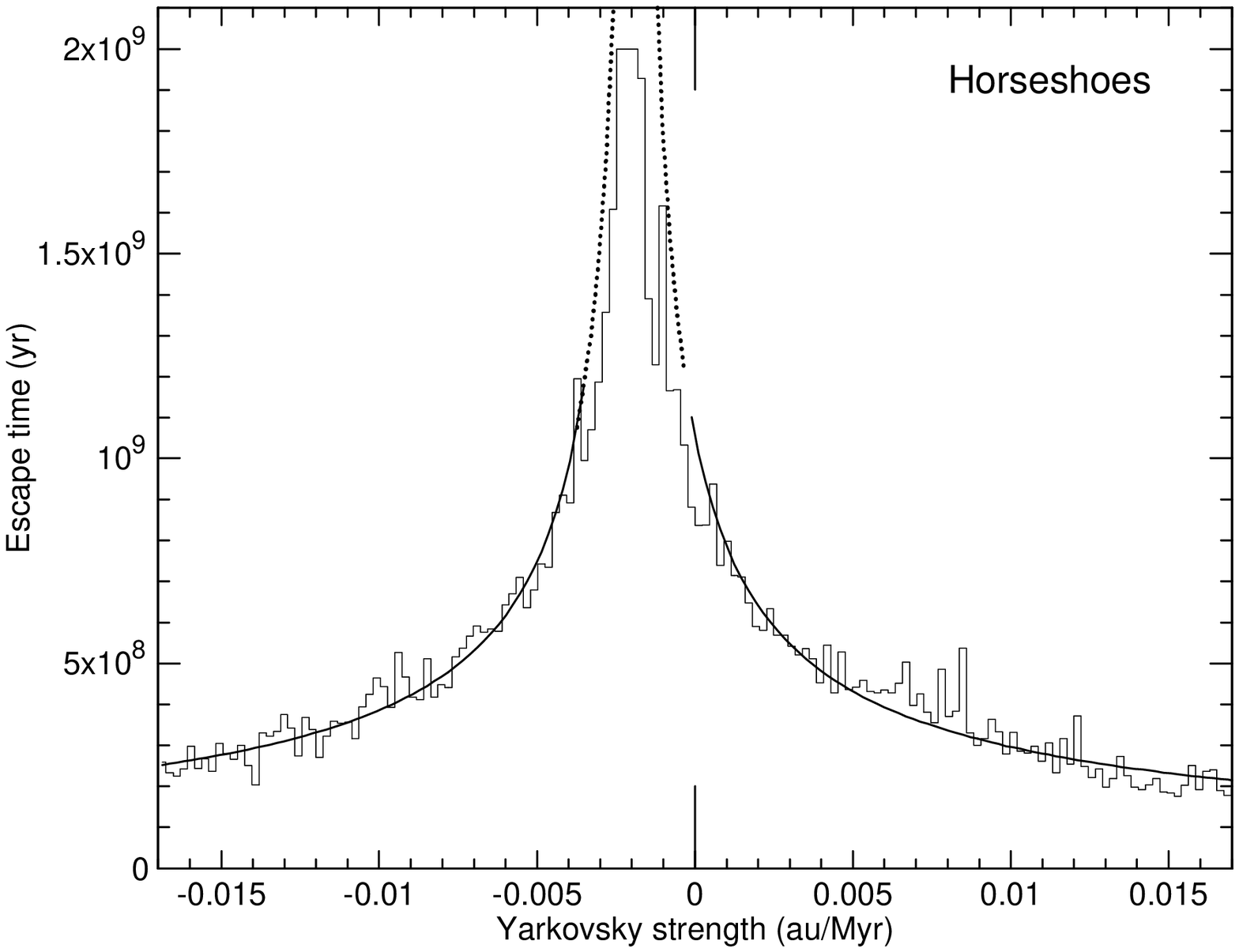}
\caption[Lifetimes of Earth tadpole (left panel) and horseshoe (right panel) co-orbitals as a function of Yarkovsky strength from the numerical simulations. The vertical line segments indicate the loci of zero Yarkovsky strength. The smooth curves represent fits to the lifetime data as described in the text.]{Lifetimes of Earth tadpole (top panel) and horseshoe (bottom panel) co-orbitals as a function of Yarkovsky strength from the numerical simulations. The vertical line segments indicate the loci of zero Yarkovsky strength. The smooth curves represent fits to the lifetime data as described in the text.}
\label{fig:lifetime_vs_yark}
\end{figure}

Co-orbital lifetime clearly depends on the magnitude of the Yarkovsky acceleration (Fig.~\ref{fig:lifetime_vs_yark}, here expressed as a constant drift rate $\dot{a}$ for the orbital semimajor axis in the absence of resonance. Interestingly, we also find that the lifetime curves for both tadpole and horseshoe particles are slightly offset with respect to the abscissa location of zero Yarkovsky strength with the effect that the longest-lived co-orbitals are those with a slightly negative value of the acceleration, corresponding to an equivalent size of $D=280$ m (tadpoles) and $D=450$ m (horseshoes). Putting this last observation aside for a moment, the anti-correlation between Yarkovsky strength and lifetime combined with the deterministic character of the trajectory (Fig~\ref{fig:l_vs_e_vs_i_yark}) suggests that orbit evolution takes place along a path in $(\Delta a, e, I)$ space that is primarily, if not uniquely, determined by initial location, say $(\Delta a_{0}, e_{0}, I_{0})$. Coorbital lifetime is then determined through the time taken to traverse the length of the path segment $\Gamma_{0,b}$ between this initial location and the point $(\Delta{a}_{b}, e_{b}, I_{b})$ where the path intersects the boundary of the stable domain. Formally, this path length is given by
\begin{equation}
S=\oint_{\Gamma_{0,b}} \hspace{-1em} ds \mbox{   where   } [ds]^2 = [d (\Delta a)]^{2} +{[d e]}^{2} + {[d I]}^{2}\mbox{.}
\end{equation}

The rate of orbit evolution $dS$/$dt$ will generally not be constant but the time average should depend monotonically on the Yarkovsky drift rate $\dot{a}$. Since $\dot{a}\propto\alpha_{Y}$ \citep{Farinella.et.al1998}, from Eq.~\ref{eq:yarko} we expect the largest asteroids to drift at the slowest pace and thus survive the longest.

Now we return to the offset for the longest-lived particles observed in Fig.~\ref{fig:lifetime_vs_yark}. In addition to our implementation of the Yarkovsky acceleration through MERCURY's user-defined force feature, we have introduced a modification to the standard MERCURY code in the form of extra housekeeping arrays to keep track of the integer identifier initially assigned to each particle within MERCURY. This is  necessary because, as particles escape from the solar system or collide with other planets or the Sun, they are removed from the simulation and their identifiers assigned to other particles. If this housekeeping does not take place as intended, one consequence would be that the Yarkovsky acceleration coefficient for a given particle is systematically shifted towards lower values by increments of $\Delta \dot{a} = 0.017/151 \simeq 10^{-4}$ au $\mbox{Myr}^{-1}$ during the simulation. To first order, we would expect this to manifest as a translation of the lifetime function along the $\dot{a}$ axis and this is, in fact, what we observe in Fig.~\ref{fig:lifetime_vs_yark}. This issue is under investigation and we will report on its resolution in a future communication. Although it may turn out that the particle handling is done correctly in the numerical code, {\it here we have chosen to treat the offset as fiducial} and taken steps to isolate it so that it does not affect our main conclusions. In what follows we describe these steps, then go on to compare our constraints of Earth co-orbital stability with those obtained in the similar investigation by \citet{Zhou.et.al2019}.

The data for the lifetime $L$ have been fitted to a function of the drift rate $y$ of the form
\begin{equation}
\label{eq:L_vs_adot}
L(y)=c {|y-y_{0}|}^{b}
\end{equation}
where we have used only data with $L < 10^{9}$ yr in the fit, in other words we use only the wings of the distributions shown in Fig.~\ref{fig:lifetime_vs_yark} where $\Delta \dot{a} \ll \dot{a}$ and ignoring the data near $\dot{a}=0$ which would be more significantly affected by the fiducials discussed in the previous paragraph. We find $\log c= 7.179 \pm 0.605$, $b=-0.643 \pm 0.285$, $y_{0}=-0.00296 \pm 0.00129$ for the tadpoles and $\log c= 7.140 \pm 0.605$, $b=-0.692 \pm 0.329$, $y_{0}=-0.00189 \pm 0.00153$ for the horseshoes. The parameter $y_{0}$ should encapsulate the bulk of any fiducial component in the data and we can simply disregard it, in effect replacing $y-y_{0}$ by $y$ in Eq.~\ref{eq:L_vs_adot}. Under these assumptions, we find that a lifetime $L = 4.5 \times 10^{9}$ yr corresponds to $\dot{a}=1.4 \times 10^{-4}$ au $\mbox{Myr}^{-1}$ for the tadpoles and $\dot{a}=2.3 \times 10^{-4}$ au $\mbox{Myr}^{-1}$ for the horseshoes, therefore asteroids with a faster Yarkovsky drift would be removed over the age of the solar system. These values can be converted to diameters using Eq.~13 of \citet{Zhou.et.al2019} with $P = 2$ hr yielding $D=1300$ m and $D=800$ m or an average of $D=1050$ m. In turn, this size translates into $H=17.1$ for $p_{V}=0.25$ typical of an S-type asteroid or $H=18.8$ for $p_{V}=0.05$ typical of C-type asteroids. For the same lifetime and albedo values, \citeauthor{Zhou.et.al2019} obtained limiting absolute magnitudes of $H=18.0$ and $H=19.7$, fainter than our estimates by 0.9 magnitudes so that the corresponding sizes are smaller by a factor of 1.6. Given that the two Yarkovsky force model implementations are generally different, the agreement between the \citeauthor{Zhou.et.al2019} constraints and this work is excellent and reinforces the effective size limits obtained for deep Earth coorbitals (but see Section~4).

\subsection{\label{sec:yplus} Outward evolution of asteroids interior to Earth's orbit}
In their work, \citet{Zhou.et.al2019} speculated that Earth's coorbital region may be preferentially populated with objects with a prograde spin and $\dot{a}>0$. This is because the Yarkovsky effect will act to reduce the libration amplitude \citep{WangHou2017} for objects with $a<1$ au and $\dot{a}>0$ arriving into the coorbital region. Here we want to find out if this is an effective pathway for an asteroid to become an Earth co-orbital and characterise the stability of these captured co-orbitals. For this purpose, we have taken 51 initial orbits from group B1, reduced their initial semimajor axis $a$ by $0.05$ au and integrated them with MERCURY for up to 300 Myr until they escaped the solar system or collided with a massive body. We consider that particles are effectively lost when either $a < 0$ (ie hyperbolic orbits) or $a > 50$ au and record the state immediately before this condition is satisfied as the final state in the dynamical evolution. We ran the same simulation setup five times, each time using a model Yarkovsky acceleration corresponding to different values of the asteroid diameter $D$: 300 m, 120 m, 47 m, 19 m and 8 m. These correspond to absolute magnitudes no brighter than $H=19-27$ with a step $\Delta H = 2$ for an albedo upper limit of $p_{V}= 0.5$. 
The equivalent Yarkovsky drift for these diameter values varies from $\dot{a}=2.8 \times 10^{-3}$ au $\mbox{Myr}^{-1}$ ($H=19$) to $\dot{a}=0.1$ au $\mbox{Myr}^{-1}$ ($H=27$). We note that, because the strength of the Yarkovsky acceleration across the 51 particles is the same in each of the runs, these simulations are not affected by the issue that may be causing fiducials in the B1Y and B4Y runs.

\begin{figure}
\hspace{-5mm}\includegraphics[width=87mm,angle=0]{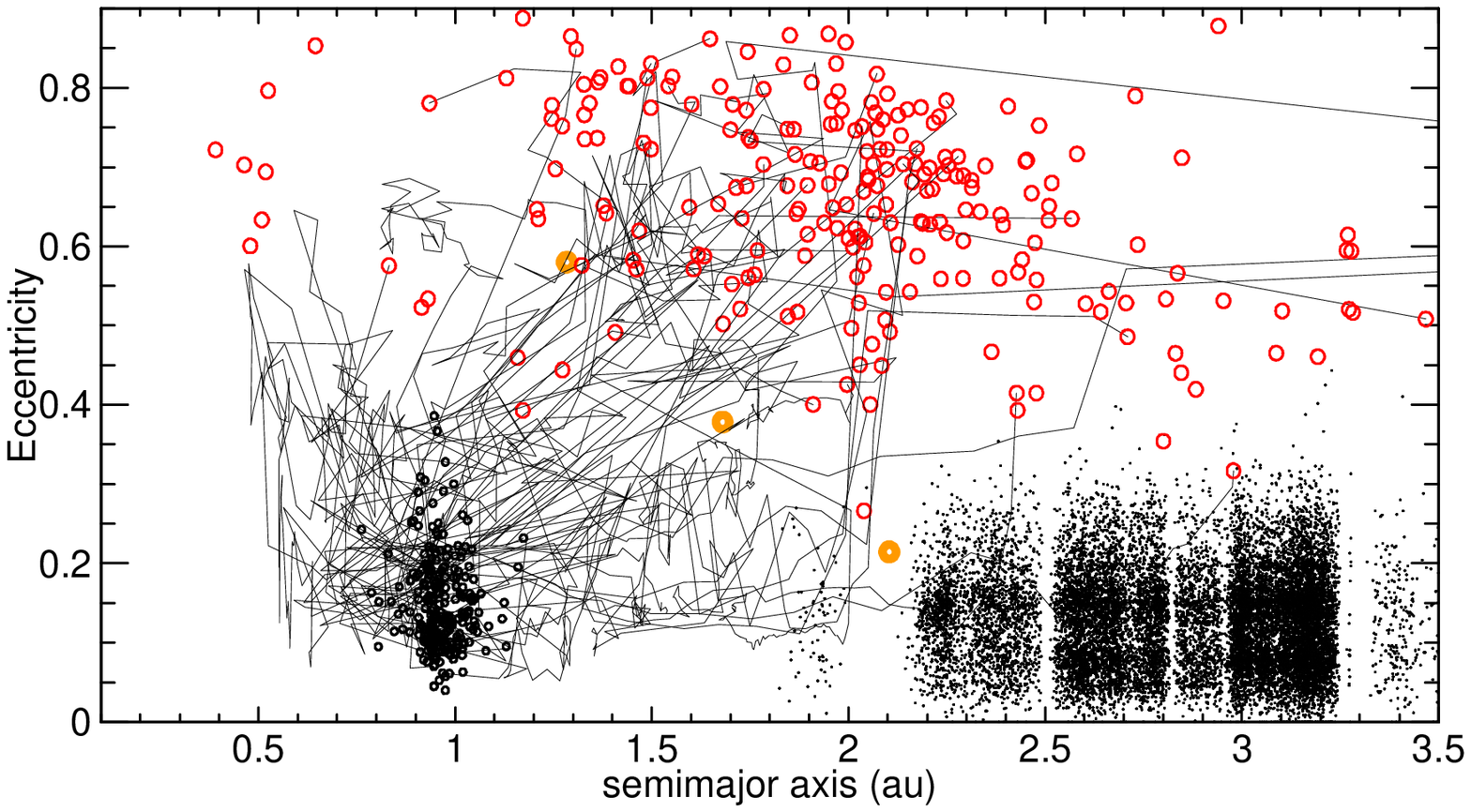}
\includegraphics[width=87mm,angle=0]{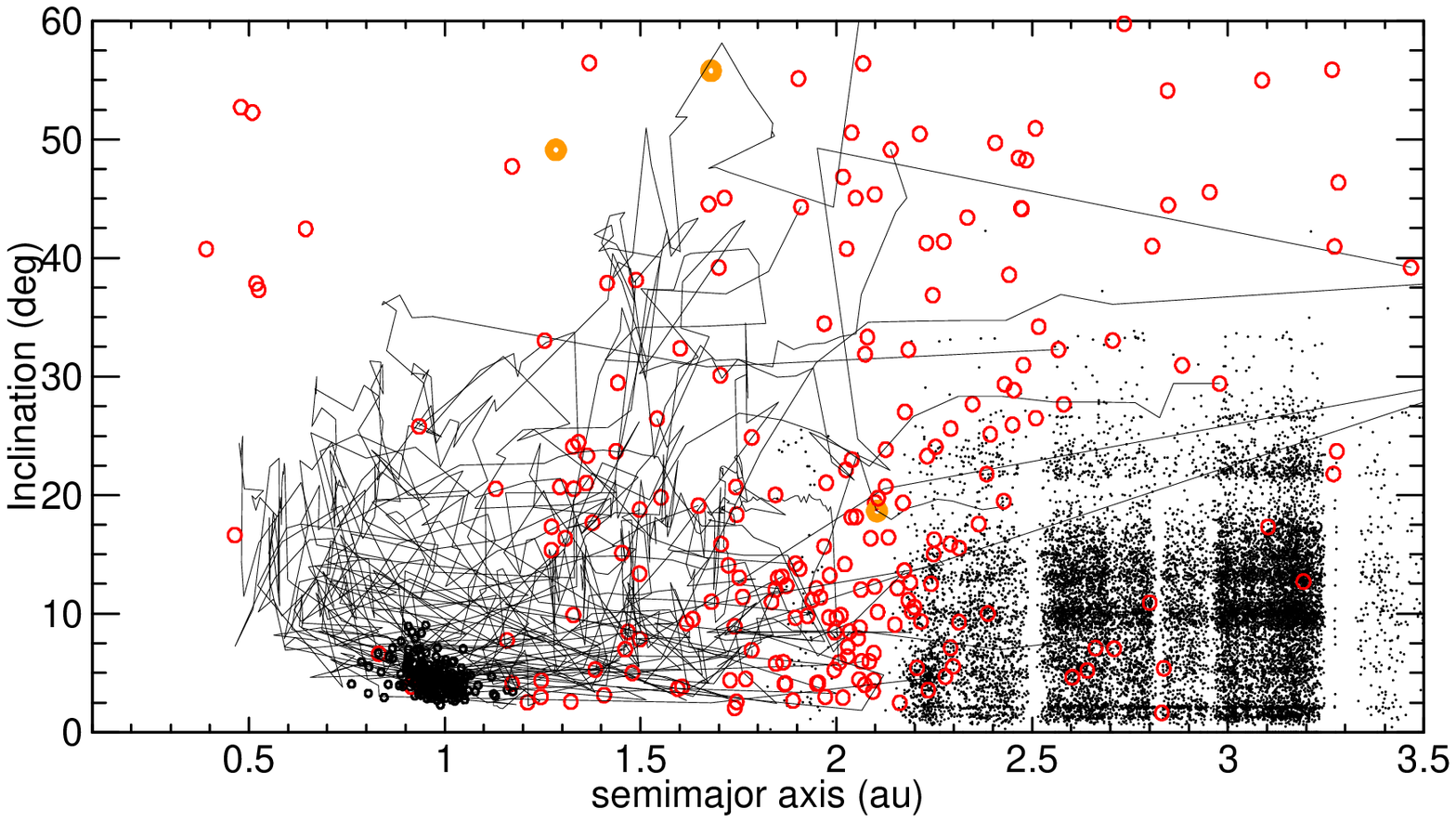}
\caption[Dynamical evolution of test particles in the C{[$H$]}) runs (Table~\ref{tab:sims}) initially at $a=0.95$ au, $e=0.025$, $I=5^{\circ}$ and with $\dot{a} > 0$. The cluster of small black circles represents the particle initial locations while the red circles indicate the location just before the particle is scattered away from the planetary region. The dots represent known the orbits of asteroids.]{Dynamical evolution of test particles in the C[$H$] runs (Table~\ref{tab:sims}) initially at $a=0.95$ au, $e=0.025$, $I=5^{\circ}$ and with $\dot{a} > 0$. The cluster of small black circles represents the particle initial locations while the red circles indicate the location just before the particle is scattered away from the planetary region. The dots represent the orbits of known asteroids.}
\label{fig:a_vs_e_vs_i_yplus}
\end{figure}

We find that particles are generally scattered away from the Earth's orbit and none become Earth co-orbitals for longer than a few million yr. Apparently, the chaotic evolution of the orbit due to Earth encounters dominates over the gradual radial drift expected by the Yarkovsky effect. Fig~\ref{fig:a_vs_e_vs_i_yplus} illustrates the routes followed by the particles from their starting location just interior to Earth's orbit (small black circles) until their final states (large red circles). The dots represent the location of Main Belt asteroids retrieved from the Asteroids Dynamic site\footnote{https://newton.spacedys.com/astdys/} where we have plotted only objects with $H<13.5$ to avoid cluttering up the plot.

In terms of the dynamical fates of the particles, we find that most are scattered onto orbits with $e$=$0.5$-$0.9$ and $a > 1$ au, following one or more moderately deep forays into the region interior to the initial orbit. The moderate clustering of escape states at a$\sim$2 au suggests that an important dynamical sink for asteroids with Earth-like orbits is the $\nu_{6}$ secular resonance at the inner edge of the Main Belt \citep{Gladman.et.al1997}. Only three particles out of a total of 256 survive in our numerical runs, these are shown as amber circles. Of those, two particles finish the simulation in high-inclination ($\gtrsim 50^{\circ}$) orbits with $e\lesssim 0.6$ in the region $1<a<2$ au and one at $a\simeq2.1$ au, $e\simeq0.2$ and $I\simeq 0^{\circ}$.

\section{\label{sec:end}Conclusions and Discussion}
The main conclusions of this paper are as follows:

We confirm earlier work \citep{Cuk.et.al2012,Zhou.et.al2019} showing that, under planetary gravitational perturbations, Earth horseshoe asteroids are significantly more stable than the Trojans that librate around the $\mbox{L}_{4}$ and $\mbox{L}_{5}$ equilibrium positions in the Sun-Earth-particle three-body problem. We further show that particles placed deep within the horseshoe region (group B4 in our simulations) remain there for at least $\sim$50\% of the age of the solar system. Our observations constrain the number of primordial co-orbital planetesimals that may have populated Earth's orbit in the past. Through a simple likelihood calculation, we show that the present absence of such planetesimals implies that typically 3-12 such objects and no more than $\sim$60 (95\% confidence) were originally present if these gradually escaped by gravitational diffusion.

By tracking the particle orbits up to the point of escaping the co-orbital resonance, we find that tadpole particles show excitation of the eccentricity and libration amplitude and leave the resonance upon reaching either $e > 0.1$ or the tadpole-horseshoe boundary at $\Delta a \simeq 0.0028$ au. In contrast, horseshoe particles escape through excitation of the inclination when $I \gtrsim 13 ^{\circ}$. Finally, co-orbitals with the highest initial libration amplitude diffuse mainly in $\Delta a$ and escape by reaching the outer boundary of the co-orbital region ($\Delta a \simeq 0.008$ au) while their eccentricity and inclination remain relatively unchanged.

These findings confirm the role of different types of secular resonances \citep{Dvorak.et.al2012,Zhou.et.al2019} in destabilising Earth co-orbital asteroids. Particle mobility in our simulations is closely correlated with the stability regions mapped out by \citet{Cuk.et.al2012} through direct numerical integration over a 3$\times$ shorter timespan, suggesting that the phase space volume occupied by long-lived orbits is not significantly eroded over the longer timescale explored in this work.

As noted in Section~\ref{sec:dearth}, The Earth exerts a certain degree of control on these long-lived co-orbitals through its own time-variable orbital eccentricity superimposed on the particles' own free eccentricity. This coupling makes it possible, in principle, to push particles out of the stable domains if $e_{\rm Earth}$ becomes too high. We have carried out a small number of simulations with slightly different initial states for the Earth and find that $e_{\rm Earth}$ remains within a narrow (of width $\sim 0.01$) range of values, therefore the likelihood of such episodes in the orbit history of the actual Earth must be small.

We then re-ran the simulations for the B1 and B4 groups with the Yarkovsky effect included in the equations of motion. We find that Yarkovsky leaves the topology of the stable domain unaltered and instead changes the character of the dynamical evolution of the co-orbitals. Particles with the same starting location evolve along a common, deterministic path until they exit the stable domain and escape. This is, in fact, similar to the mechanism leading to the escape of Mars Trojans \citep{Christou.et.al2020} with the difference that the domain of stability for Mars Trojans is located at moderate inclination \cite[$\gtrsim 13^{\circ}$;][]{Scholl.et.al2005}. The orbit evolution of the co-orbitals under Yarkovsky is more rapid than in the gravitational runs. Using group B4 as a reference, we find that particles with the highest Yarkovsky drift rate ($\sim$$ 17 \times 10^{-3}$ au $\mbox{Myr}^{-1}$) escaped after $\sim 2 \times 10^{8}$ yr and all but three particles escaped before the 2 Gyr mark. Escape time is inversely correlated with the magnitude of $\dot{a}$, as also found by \citet{Zhou.et.al2019}.

Finally, we have examined the \citeauthor{Zhou.et.al2019} conjecture that the Yarkosvky effect could be efficienty populating the Earth's coorbital region with prograde-spinning asteroids. We find that asteroids with orbits initially with $a = 0.95$ au and $\dot{a}= +0.0028\cdots +0.1$ au $\mbox{Myr}^{-1}$ are rapidly (on a timescale $<< \Delta a / \dot{a}$) scattered far from the Earth's orbit. Due to our limited sampling of phase space, it is possible that different starting orbits eg at higher inclination may produce a qualitatively different outcome. However, our results do suggest that long-lived Earth co-orbital asteroids in low-$e$, low-$I$ orbits are unlikely to have a source region interior to the Earth's orbit.

Implicit in our conclusions about the destabilising action of the Yarkovsky effect is the assumption that these asteroids existed as independent objects since their deposition at 1 au. This need not, however, be the case and the problem of retainment becomes less acute if these asteroids were derived from collisional or rotational disruption of larger objects, less affected by non-gravitational forces. Although no formal estimates of collisional lifetime exist for Earth Trojans, \citet{Bottke.et.al1994} calculated collisional lifetimes for NEAs against different impactor types and we would argue that their figures for NEA-NEA collisions are appropriate for objects in near-planar, near-circular orbits at $a\sim1$ au. Taking $D=10$ km as a reference size for a parent body, from Figures 4 \& 5 of \citeauthor{Bottke.et.al1994}, we see that such an object suffers a catastrophic impact with another NEA once every $\sim$5 Gyr if some of the impact energy goes into overcoming a non-zero internal strength for the body, or once every $\gtrsim 1$ Gyr if the target is a strengthless ``rubble pile''.

A process that would also act to extend the dynamical lifetime of Earth co-orbitals is random re-orientation by collisions or by YORP evolution. The spin axes of Main Belt asteroids are reset by rotational evolution under the YORP effect on timescales of $10^{6}-10^{7}$ yr \citep{Jacobson.et.al2014} acting to reverse the sign of $\dot{a}$, the so-called stochastic YORP model \cite[see][and references therein]{Vokrouhlicky.et.al2017} therefore on Gyr timescales the orbital evolution of asteroids due to Yarkovsky will generally be slower than if a constant drift rate was applied continuously. This will also be true for asteroids at 1 au, all the more so due to the inverse-square semimajor axis dependence of the YORP evolution timescale \citep{Jacobson.et.al2014}. Quantitative assessment of these scenaria will be the subject of future work.

\section*{Acknowledgements}
Work by AAC was supported via grant ST/R000573/1 from the UK Science and Technology Facilities Council (STFC). We thank Matija \'{C}uk for sharing his data on stable Earth coorbitals. We acknowledge the SFI/HEA Irish Centre for High-End Computing (ICHEC), the Dublin Institute for Advanced Studies (DIAS) as well as the University of Florida (UF) Department of Astronomy for the provision of computational facilities and support. We would like to thank the High Performance Computing Resources team at New York University Abu Dhabi and especially Jorge Naranjo for helping us with our numerical simulations. Astronomical research at the Armagh Observatory and Planetarium is grant-aided by the Northern Ireland Department for Communities (DfC).


\section*{Data Availability}
The data underlying this paper were accessed from the Near Earth Objects Dynamic site (https://newton.spacedys.com/neodys/), the Asteroids Dynamic site (https://newton.spacedys.com/{\linebreak}astdys/) and the JPL HORIZONS ephemeris service (https://ssd.jpl.nasa.gov/?horizons\#telnet). The derived data generated in this research are available from the corresponding author upon reasonable request.




\bibliographystyle{mnras}
\bibliography{cg_mnras_2021} 

\begin{thebibliography}{}
\makeatletter
\relax
\def\mn@urlcharsother{\let\do\@makeother \do\$\do\&\do\#\do\^\do\_\do\%\do\~}
\def\mn@doi{\begingroup\mn@urlcharsother \@ifnextchar [ {\mn@doi@}
  {\mn@doi@[]}}
\def\mn@doi@[#1]#2{\def\@tempa{#1}\ifx\@tempa\@empty \href
  {http://dx.doi.org/#2} {doi:#2}\else \href {http://dx.doi.org/#2} {#1}\fi
  \endgroup}
\def\mn@eprint#1#2{\mn@eprint@#1:#2::\@nil}
\def\mn@eprint@arXiv#1{\href {http://arxiv.org/abs/#1} {{\tt arXiv:#1}}}
\def\mn@eprint@dblp#1{\href {http://dblp.uni-trier.de/rec/bibtex/#1.xml}
  {dblp:#1}}
\def\mn@eprint@#1:#2:#3:#4\@nil{\def\@tempa {#1}\def\@tempb {#2}\def\@tempc
  {#3}\ifx \@tempc \@empty \let \@tempc \@tempb \let \@tempb \@tempa \fi \ifx
  \@tempb \@empty \def\@tempb {arXiv}\fi \@ifundefined
  {mn@eprint@\@tempb}{\@tempb:\@tempc}{\expandafter \expandafter \csname
  mn@eprint@\@tempb\endcsname \expandafter{\@tempc}}}

\bibitem[\protect\citeauthoryear{{Bottke}, {Nolan}, {Greenberg}  \&
  {Kolvoord}}{{Bottke} et~al.}{1994}]{Bottke.et.al1994}
{Bottke} W.~F.,  {Nolan} M.~C.,  {Greenberg} R.,   {Kolvoord} R.~A.,  1994, in
  Hazards Due to Comets and Asteroids. University of Arizona Press, pp 337--357

\bibitem[\protect\citeauthoryear{{Bottke}, {Durda}, {Nesvorn\'{y}}, {Jedicke},
  {Morbidelli}, {Vokrouhlick\'{y}}  \& {Levison}}{{Bottke}
  et~al.}{2005}]{Bottke.et.al2005}
{Bottke} W.~F.,  {Durda} D.~D.,  {Nesvorn\'{y}} D.,  {Jedicke} R.,
  {Morbidelli} A.,  {Vokrouhlick\'{y}} D.,   {Levison} H.~F.,  2005, Icarus,
  175, 111

\bibitem[\protect\citeauthoryear{{Bottke}, {Vokrouhlick\'{y}}, {Rubincam}  \&
  {Nesvorn\'{y}}}{{Bottke} et~al.}{2006}]{Bottke.et.al2006}
{Bottke} W.~F.,  {Vokrouhlick\'{y}} D.,  {Rubincam} D.~P.,   {Nesvorn\'{y}} D.,
   2006, Ann.~Rev.~Earth Planet.~Sci., 34, 157

\bibitem[\protect\citeauthoryear{{Bottke}, {Levison}, {Nesvorn\'{y}}  \&
  {Dones}}{{Bottke} et~al.}{2007}]{Bottke.et.al2007b}
{Bottke} W.~F.,  {Levison} H.~F.,  {Nesvorn\'{y}} D.,   {Dones} D.,  2007,
  Icarus, 190, 203

\bibitem[\protect\citeauthoryear{{Bottke}, {Walker}, {Day}, {Nesvorn\'{y}}  \&
  {Elkins-Tanton}}{{Bottke} et~al.}{2010}]{Bottke.et.al2010}
{Bottke} W.~F.,  {Walker} R.~J.,  {Day} J.~M.~D.,  {Nesvorn\'{y}} D.,
  {Elkins-Tanton} L.,  2010, Science, 330, 1527

\bibitem[\protect\citeauthoryear{{Bottke}, {Bro\v{z}}, {O'Brien}, {Campo
  Bagatin}, {Morbidelli}  \& {Marchi}}{{Bottke}
  et~al.}{2015}]{Bottke.et.al2015}
{Bottke} W.~F.,  {Bro\v{z}} M.,  {O'Brien} D.~P.,  {Campo Bagatin} A.,
  {Morbidelli} A.,   {Marchi} S.,  2015, In: Asteroids IV (P.~Michel, F.~E.
  DeMeo, W.~F. Bottke Jr.), Arizona University Press, Tucson, pp 701--724

\bibitem[\protect\citeauthoryear{{Cambioni} et~al.,}{{Cambioni}
  et~al.}{2018}]{Cambioni.et.al2018}
{Cambioni} S.,  et~al., 2018, LPI Contr.~2083, p.~1149

\bibitem[\protect\citeauthoryear{{Chambers}}{{Chambers}}{1999}]{Chambers1999}
{Chambers} J.~E.,  1999, MNRAS, 304, 793

\bibitem[\protect\citeauthoryear{{Christou}}{{Christou}}{2000}]{Christou2000a}
{Christou} A.~A.,  2000, Icarus, 144, 1

\bibitem[\protect\citeauthoryear{{Christou}}{{Christou}}{2013}]{Christou2013}
{Christou} A.~A.,  2013, Icarus, 224, 144

\bibitem[\protect\citeauthoryear{{Christou} \& {Asher}}{{Christou} \&
  {Asher}}{2011}]{ChristouAsher2011}
{Christou} A.~A.,  {Asher} D.~J.,  2011, MNRAS, 414, 2965

\bibitem[\protect\citeauthoryear{{Christou}, {Borisov}, {Dell'Oro}, {Jacobson},
  {Cellino}  \& {Unda-Sanzana}}{{Christou} et~al.}{2020}]{Christou.et.al2020}
{Christou} A.~A.,  {Borisov} G.~B.,  {Dell'Oro} A.,  {Jacobson} S.~A.,
  {Cellino} A.,   {Unda-Sanzana} E.,  2020, Icarus, 335, 113370

\bibitem[\protect\citeauthoryear{{Connors}, {Veillet}, {Brasser}, {Wiegert},
  {Chodas}, {Mikkola}  \& {Innanen}}{{Connors}
  et~al.}{2004}]{Connors.et.al2004}
{Connors} M.,  {Veillet} C.,  {Brasser} R.,  {Wiegert} P.,  {Chodas} P.,
  {Mikkola} S.,   {Innanen} K.,  2004, Met.~Planet.~Sci., 39, 1251

\bibitem[\protect\citeauthoryear{{Connors}, {Wiegert}  \& {Veillet}}{{Connors}
  et~al.}{2011}]{Connors.et.al2011}
{Connors} M.,  {Wiegert} P.,   {Veillet} C.,  2011, Nature, 475, 481

\bibitem[\protect\citeauthoryear{{\'{C}uk}, {Hamilton}  \& {Holman}}{{\'{C}uk}
  et~al.}{2012}]{Cuk.et.al2012}
{\'{C}uk} M.,  {Hamilton} D.~P.,   {Holman} M.~J.,  2012, MNRAS, 426, 3051

\bibitem[\protect\citeauthoryear{{\'{C}uk}, {Christou}  \&
  {Hamilton}}{{\'{C}uk} et~al.}{2015}]{Cuk.et.al2015}
{\'{C}uk} M.,  {Christou} A.~A.,   {Hamilton} D.~P.,  2015, Icarus, 252, 339

\bibitem[\protect\citeauthoryear{{Dohnanyi}}{{Dohnanyi}}{1969}]{Dohnanyi1969}
{Dohnanyi} J.~S.,  1969, J.~Geophys.~Res., 74, 2531

\bibitem[\protect\citeauthoryear{{Duddy}, {Lowry}, {Wolters}, {Christou},
  {Weissman}, {Green}  \& {Rozitis}}{{Duddy} et~al.}{2012}]{Duddy.et.al2012}
{Duddy} S.~R.,  {Lowry} S.~C.,  {Wolters} S.~D.,  {Christou} A.~A.,  {Weissman}
  P.,  {Green} S.~F.,   {Rozitis} B.,  2012, A\&A, 539, A36

\bibitem[\protect\citeauthoryear{{Dvorak}, {Lhotka}  \& {Zhou}}{{Dvorak}
  et~al.}{2012}]{Dvorak.et.al2012}
{Dvorak} R.,  {Lhotka} C.,   {Zhou} L.,  2012, A\&A, 541, A127

\bibitem[\protect\citeauthoryear{{Farinella}, {Vokrouhlick\'{y}}  \&
  {Hartmann}}{{Farinella} et~al.}{1998}]{Farinella.et.al1998}
{Farinella} P.,  {Vokrouhlick\'{y}} D.,   {Hartmann} W.~D.,  1998, Icarus, 132,
  378

\bibitem[\protect\citeauthoryear{{Georgakarakos}, {Dobbs-Dixon}  \&
  {Way}}{{Georgakarakos} et~al.}{2016}]{Georgakarakos.et.al2016}
{Georgakarakos} N.,  {Dobbs-Dixon} I.,   {Way} M.~J.,  2016, MNRAS, 461, 1512

\bibitem[\protect\citeauthoryear{{Giorgini} et~al.,}{{Giorgini}
  et~al.}{1996}]{Giorgini.et.al1996}
{Giorgini} J.~D.,  et~al., 1996, BAAS, 28, 1158

\bibitem[\protect\citeauthoryear{{Gladman} et~al.,}{{Gladman}
  et~al.}{1997}]{Gladman.et.al1997}
{Gladman} B.~J.,  et~al., 1997, Science, 277, 197

\bibitem[\protect\citeauthoryear{{Jacobson}, {Marzari}, {Rossi}, {Scheeres}  \&
  {Davis}}{{Jacobson} et~al.}{2014}]{Jacobson.et.al2014}
{Jacobson} S.~A.,  {Marzari} F.,  {Rossi} A.,  {Scheeres} D.~J.,   {Davis}
  D.~R.,  2014, MNRAS, 439, L95

\bibitem[\protect\citeauthoryear{{Laskar}}{{Laskar}}{1994}]{Laskar1994}
{Laskar} J.,  1994, A\&A, 287, L9

\bibitem[\protect\citeauthoryear{{Malhotra}}{{Malhotra}}{2019}]{Malhotra2019}
{Malhotra} R.,  2019, Nature Astronomy, 3, 193

\bibitem[\protect\citeauthoryear{{Markwardt}, {Gerdes}, {Malhotra}, {Becker},
  {Hamilton}  \& {Adams}}{{Markwardt} et~al.}{2020}]{Markwardt.et.al2020}
{Markwardt} L.,  {Gerdes} D.~W.,  {Malhotra} R.,  {Becker} J.~C.,  {Hamilton}
  S.~J.,   {Adams} F.~C.,  2020, MNRAS, 492, 6105

\bibitem[\protect\citeauthoryear{{Marzari} \& {Scholl}}{{Marzari} \&
  {Scholl}}{2013}]{MarzariScholl2013}
{Marzari} F.,  {Scholl} H.,  2013, Cel.~Mech.~Dyn.~Astron., 117, 91

\bibitem[\protect\citeauthoryear{{Morais}}{{Morais}}{1999}]{Morais1999}
{Morais} M.~H.~M.,  1999, A\&A, 350, 318

\bibitem[\protect\citeauthoryear{{Morais} \& {Morbidelli}}{{Morais} \&
  {Morbidelli}}{2002}]{MoraisMorbidelli2002}
{Morais} M.~H.~M.,  {Morbidelli} A.,  2002, Icarus, 160, 1

\bibitem[\protect\citeauthoryear{{Murray} \& {Dermott}}{{Murray} \&
  {Dermott}}{1999}]{MurrayDermott1999}
{Murray} C.~D.,  {Dermott} S.~F.,  1999, {Solar System Dynamics}.
Cambridge University Press, Cambridge

\bibitem[\protect\citeauthoryear{{Scholl}, {Marzari}  \& {Tricarico}}{{Scholl}
  et~al.}{2005}]{Scholl.et.al2005}
{Scholl} H.,  {Marzari} F.,   {Tricarico} P.,  2005, Icarus, 175, 397

\bibitem[\protect\citeauthoryear{{Tabachnik} \& {Evans}}{{Tabachnik} \&
  {Evans}}{2000}]{TabachnikEvans2000a}
{Tabachnik} S.,  {Evans} N.~W.,  2000, MNRAS, 319, 63

\bibitem[\protect\citeauthoryear{{Tsiganis}, {Dvorak}  \&
  {Pilat-Loginger}}{{Tsiganis} et~al.}{2000}]{Tsiganis.et.al2000}
{Tsiganis} K.,  {Dvorak} R.,   {Pilat-Loginger} E.,  2000, A\&A, 354, 1091

\bibitem[\protect\citeauthoryear{{Vokrouhlick\'{y}}, {Bottke}  \&
  {Nesvorn\'{y}}}{{Vokrouhlick\'{y}} et~al.}{2017}]{Vokrouhlicky.et.al2017}
{Vokrouhlick\'{y}} D.,  {Bottke} W.~F.,   {Nesvorn\'{y}} D.,  2017, AJ, 153,
  id.~172

\bibitem[\protect\citeauthoryear{{Wang} \& {Hou}}{{Wang} \&
  {Hou}}{2017}]{WangHou2017}
{Wang} X.,  {Hou} X.,  2017, MNRAS, 471, 243

\bibitem[\protect\citeauthoryear{{Weidenschilling}}{{Weidenschilling}}{1977}]{Weidenschilling1977}
{Weidenschilling} S.~J.,  1977, Astrophys.~Space Sci., 51, 153

\bibitem[\protect\citeauthoryear{{Weissman} \& {Wetherill}}{{Weissman} \&
  {Wetherill}}{1974}]{WeissmanWetherill1974}
{Weissman} P.~R.,  {Wetherill} G.~W.,  1974, AJ, 79, 404

\bibitem[\protect\citeauthoryear{{Xu}, {Zhou}, {Dvorak}  \& {Ip}}{{Xu}
  et~al.}{2020}]{Xu.et.al2020}
{Xu} Y.-B.,  {Zhou} L.-Y.,  {Dvorak} R.,   {Ip} W.-H.,  2020, MNRAS, 493, 1447

\bibitem[\protect\citeauthoryear{{Yoshikawa}, {Tsuda}, {Watanabe}, {Tanaka},
  {Nakazawa}, {Terui}  \& {Saiki}}{{Yoshikawa}
  et~al.}{2018}]{Yoshikawa.et.al2018}
{Yoshikawa} M.,  {Tsuda} Y.,  {Watanabe} S.,  {Tanaka} S.,  {Nakazawa} S.,
  {Terui} F.,   {Saiki} T.,  2018, LPI Contr.~2083, p.~1771

\bibitem[\protect\citeauthoryear{{Zhou}, {Xu}, {Zhou}, {Dvorak}  \&
  {Li}}{{Zhou} et~al.}{2019}]{Zhou.et.al2019}
{Zhou} L.,  {Xu} Y.-B.,  {Zhou} L.-Y.,  {Dvorak} R.,   {Li} J.,  2019, A\&A,
  622, A97

\makeatother
\end{thebibliography}
\clearpage






\begin{table*}
\centering
\caption[Summary of the different groups of simulation runs.]{Summary of the different groups of simulation runs.}
\begin{tabular}{lccc}
\noalign{\smallskip}
\hline \hline
Group  & Type of & \# of groups $\times$ & Duration \\
ID  & Co-orbital & \# of particles/group & (Gyr) \\\hline
\multicolumn{4}{c}{Gravity, 8 Planets} \\\hline
B[$i$], $i=1,\cdots,6$ & 2$\times$tadpole, 4$\times$horseshoe & 6$\times$151 & 2 Gyr \\
B1E[$j$], B4E[$j$], $j=1,\cdots,3$ & 3$\times$tadpole, 3$\times$horseshoe & 3$\times$ 2$\times$51 & 0.9 Gyr \\\hline
\multicolumn{4}{c}{Gravity, 8 Planets + Yarkovsky} \\\hline
B1Y, B4Y & 1$\times$tadpole, 1$\times$horseshoe & 2$\times$151 & 2 Gyr \\
C[$H$], $H=19,21,\cdots,27$ & $a=0.95$ au & {\bf 5}$\times$51 & 100-300 Myr  \\
\hline \hline
\end{tabular}
\label{tab:sims}
\end{table*}
\clearpage

\begin{table*}
\centering
\caption[Initial conditions for the test particles in the numerical simulations${}^{\dagger}$]{Initial conditions for the test particles in the numerical simulations${}^{\dagger}$}
\begin{tabular}{lcccccc}
\noalign{\smallskip}
\hline \hline
Group &  $a$-$a_{\rm Earth}$   &    & $I$  & $\omega$ & $\Omega$ & M \\
ID & (au)      & $e$ & (deg)  & (deg) & (deg) & (deg) \\ \hline
\blap{B1\\B2\\B3\\B4\\B5\\B6}  & \blap{$0.00097268$\\$0.00249256$\\$0.00406922$\\$0.00578675$\\$0.00741953$\\$0.00905588$}  &  &  &  &  &  \\
& &  &  &  &  &  \\
&  &    $0.025$   & $5.0$   & $108.542$ & $40.4758$ & $148.969$ \\
& &  &  &  &  &  \\
& &  &  &  &  &  \\
& &  &  &  &  &  \\
\hline \hline
\multicolumn{7}{l}{\parbox{117mm}{${}^{\dagger}$ At the epoch JD 2456190.5 = 20 September 2012, 0.0 UT.}}
\end{tabular}
\label{tab:ic}
\end{table*}
\clearpage

\begin{table}
\centering
\caption[Summary of particle fates in the N-body simulations.]{Summary of particle fates in the N-body simulations.}
\begin{tabular}{lccc}
\noalign{\smallskip}
\hline \hline
Group  & Total \# & \# escaped & \# remaining\\
ID  & of particles & after 2 Gyr & after 2 Gyr \\\hline
B1 & 151 & 125 (83\%) & 26 (17\%) \\
B2 & 151 & 150 (99\%) & 1 (1\%)\\ \hline
Tadpoles & 302 & 275 (91\%) & 27 (9\%) \\
\hline
B3 & 141 & 121 (86\%) & 20 (14\%) \\
B4 & 143 & 0 (0\%) & 143 (100\%)\\
B5 & 151 & 78 (52\%) & 73 (48\%)\\  \hline
Horseshoes & 435 & 199 (46\%)  & 236  (54\%)\\
\hline
B6 & 149 & 149 (100\%) & 0 (0\%)\\ \hline \hline
\end{tabular}
\label{tab:esc}
\end{table}
\clearpage

\clearpage

\clearpage

\clearpage

\clearpage

\clearpage

\clearpage

\clearpage

\clearpage

\bsp	
\label{lastpage}
\end{document}